# Early-career setback and future career impact

Yang Wang [1,2], Benjamin F. Jones [1,2,3], and Dashun Wang [1,2,4] [*]

[1] *Northwestern Institute on Complex Systems, Northwestern University, Evanston, IL, USA*

[2] *Kellogg School of Management, Northwestern University, Evanston, IL, USA*

[3] *National Bureau of Economic Research (NBER), Cambridge, MA, USA*

[4] *McCormick School of Engineering, Northwestern University, Evanston, IL, USA*

[*] To whom correspondence should be addressed. e-mail: dashun.wang@northwestern.edu

# ABSTRACT

Setbacks are an integral part of a scientific career, yet little is known about whether an early-career setback may augment or hamper an individual's future career impact. Here we examine junior scientists applying for U.S. National Institutes of Health (NIH) R01 grants. By focusing on grant proposals that fell just below and just above the funding threshold, we compare "near-miss" with "near-win" individuals to examine longer-term career outcomes. Our analyses reveal that an early-career near miss has powerful, opposing effects. On one hand, it significantly increases attrition, with one near miss predicting more than a 10% chance of disappearing permanently from the NIH system. Yet, despite an early setback, individuals with near misses systematically outperformed those with near wins in the longer run, as their publications in the next ten years garnered substantially higher impact. We further find that this performance advantage seems to go beyond a screening mechanism, whereby a more selected fraction of near-miss applicants remained than the near winners, suggesting that early-career setback appears to cause a performance improvement among those who persevere. Overall, the findings are consistent with the concept that "what doesn't kill me makes me stronger." Whereas science is often viewed as a setting where early success begets future success, our findings unveil an intimate yet previously unknown relationship where early-career setback can become a marker for future achievement, which may have broad implications for identifying, training and nurturing junior scientists whose career will have lasting impact.

"Science is 99 percent failure, and that's an optimistic view", said Robert Lefkowitz, who was awarded the Nobel prize in 2012 for his groundbreaking studies of G-protein-coupled receptors. Despite the ubiquitous nature of failures, it remains unclear if a setback in an early career may augment or hamper an individual's future career impact. Indeed, the Matthew effect[1-9] suggests a "rich get richer" phenomenon where early-career success helps bring future victories. In addition to community recognition, bringing future attention and resources[5,7,8,10-14], success may also influence individual motivation[15], where positive feedback bolsters self-confidence. Together, these views indicate that it is early-career success, not failure, that would lead to future success. Yet at the same time other mechanisms suggest that the opposite may also be true. Indeed, screening mechanisms[16,17] suggest that, if early-career failures screen out less-determined researchers, early setbacks among those who remain could, perhaps counterintuitively, become a marker for future achievement. Further, failure may teach valuable lessons that are hard to learn otherwise[18-20], while also motivating individuals to redouble effort[21,22]. Such positive views of failure are reflected in Nietzsche's classic phrase "what doesn't kill me makes me stronger"[23], in the celebration-of-failure mindset in Silicon Valley[24], and in a recent commencement address by U.S. Supreme Court chief justice John Roberts, who told graduating students "I wish you bad luck." Overall, these divergent perspectives indicate that the net effect of an early-career setback is unclear. Given the consequential nature of this question to individual careers and the institutions that support and nurture them, and building on the remarkable progress in our quantitative understanding of science[1,2,7-9,25-49], here we ask: Can an early-career setback lead to future career impact?

To offer quantitative answers to this question, we leveraged a unique dataset, containing all R01 grant applications ever submitted to the NIH, to examine early-career success and failure. NIH funding decisions are largely determined by paylines derived from evaluation scores. Our empirical strategy harnesses the highly nonlinear relationship between funding success and evaluation score around the funding threshold (Fig. 1a). Indeed, focusing on individuals whose proposals fell just above and below the threshold allows us to compare observationally-similar individuals who are either "near misses" (individuals who just missed receiving funding) or "near winners" (individuals who just succeeded in getting funded). Here, we focus on junior scientists by examining principal investigators (PIs) whose first application to the NIH was within the previous three years. We combined the NIH grant database with the Web of Science data, tracing all competing R01 grant applications submitted to the NIH between 1990 and 2005 together with research outputs by the PIs, measured by their publication and citation records (SI Appendix S1).

We examined performance and demographic characteristics for the two groups of PIs, finding that prior to treatment, they are statistically indistinguishable along all dimensions we measured (Fig. 1b). Yet the treatment created a clear difference between the two, whereby one group was awarded R01 grants, which on average amount to $1.3 million for five years, while the other group was not. Given the pre-treatment similarity between the two groups, we ask: which group produced works with higher impacts over subsequent years?

**Results**

We therefore traced the publication records of PIs from the two groups. We first focus on active PIs in the NIH system, defined as those who apply for and/or receive NIH grants at some point in the future (Fig. 1c and SI Appendix S3.9). We calculated the publication rates of the PIs, finding that the two groups published a similar number of papers per person over the next ten-year period (Fig. 2a), consistent with prior studies[8,50]. We then computed, out of the papers published by the near-miss and near-win group, the probability of finding hit papers (Fig. 2b), defined as being in the top 5% of citations received in the same field and year[32,35]. In the first five years, 13.3% of papers published by the near-win group turned out to be a hit paper, which is substantially higher than the baseline hit rate of 5%, demonstrating that near winners considered in our sample produced hit papers at a much higher rate than average scientists in their field. We measured the same probability for the near-miss group, finding that they produced hit papers at an average rate of 16.1%. Comparing the two groups, we find near misses outperformed near winners significantly, by a factor of 21% ($\chi^2$-test $p$-value < 0.001). This performance advantage persisted: We analyzed papers produced in the second five-year window (6-10 years after treatment), uncovering a similar gap (Fig. 2b). To ensure the observed effect is not just limited to hit papers, we also quantified performance using other commonly used measures, including average citations received within five years of publication (Fig. 2c) and the relative citation ratio (RCR) of each paper (SI Appendix S3.4)[51], arriving at the same conclusions. Indeed, papers published by the near-miss group in the next two five-year periods attracted on average 18.8% and 11% more citations than those by the near-win group, respectively (Fig. 2c, *p*-value < 0.001).

To further test the robustness of our results we repeated our analyses along several dimensions. We changed our definitions of junior PIs to two alternatives, by focusing on first-time R01 applicants only, and by restricting to those without any current NIH grants (SI Appendix S3.2). We varied our definitions of hit papers (from top 1% to top 15% of citations, SI AppendixS3.3). We also computed per capita measures of hit papers (SI Appendix S3.1). We further adjusted for field differences of citations, by calculating the average normalized citations by field and year[32] (SI Appendix S3.4). We also repeated our analyses across different measurement time periods (SI Appendix S3.5). Amid all variations, the conclusions remain the same.

The performance advantage of the near misses is particularly surprising given that the near winners, by construction, had an initial NIH funding advantage immediately after treatment. Given that funding from the NIH can be an important means to augmenting scientific production[33,36,37,50,52-54], we investigate funding dynamics for the near-miss and near-win groups over the following ten-year period. We find that the near-miss group naturally received significantly less NIH funding in the first five years following treatment, averaging $0.27 million less per person (Fig. 2d, *p*-value < 0.05), which is consistent with prior studies[8,50,55]. Yet the funding difference between the two groups disappeared in the second five-year period (Fig. 2d, *p*-value > 0.1). Note that, although NIH is the world's largest public funder for biomedical research, near misses may have received more funding elsewhere. While we cannot test all potential sources of funding, we further measured the total funding support from the U.S. National Science Foundation (NSF) received by individuals with the same name in the same period, finding a lack of difference between the two groups (SI Appendix S3.6, Fig. S15). The NIH and NSF together account for the vast majority of funding obtained by junior biomedical scientists in the US (SI Appendix S3.6)[33].

Together, these results demonstrate that over the course of ten years, near misses had fewer initial grants from the NIH. Yet they ultimately published as many papers and, most surprisingly, produced work that garnered substantially higher impacts than their near-win counterparts.

Is the uncovered difference in outcomes causally attributable to the early-career setback? Or, could it be explained by other alternative forces? Indeed, there might still exist unobserved influences on funding success near the threshold (e.g., individual characteristics, fields of study, personality traits, etc.), which might also drive future career outcomes. To rule out alternative explanations, we leverage the causal inference technique fuzzy Regression Discontinuity (RD)[56,57]. Specifically, we use an indicator for the score being above or below the funding threshold as an instrumental variable (IV), rather than the actual funding outcome itself, to predict future career outcomes (see Methods section). The RD approach helps us rule out unobserved influences on funding outcome or any otherwise unobserved individual characteristics that differ smoothly with the score[56,57], allowing us to further establish a causal link between early-career near miss and future career impact. By accounting for any potential confounding factors, our RD estimates indicate that one early-career near miss increases the probability of publishing a hit paper in the next 10 years by 6.1% (*p*-value 0.041), and the average citations per paper by 34% (9.67 citations in 5 years, *p*-value 0.046) (see Methods section). Taken together, the RD analyses establish the causal interpretation of our results, and its agreement with the near-win versus near-miss approach used in Fig. 2 further demonstrates the robustness of our findings.

These results document that, despite an early setback, near misses outperformed near winners over the longer run, conditional on remaining active in the NIH system. This finding has a striking implication. Namely, take two researchers who are seeking to continue their careers in science. *All else equal*, except that one has had an early funding failure and the other an early funding success, it is the one who failed that is expected to write higher-impact papers in the future.

To conceptualize this finding, consider two hypotheses. The first is a screening hypothesis, where the population of "survivors" among the near-miss group may have fixed, advantageous characteristics. Second, the result is consistent with failure itself teaching valuable lessons or strengthening resolve. To help unpack the findings, we examine differential survival rates between two samples and further ask whether the screening hypothesis alone may be sufficient to explain the observed difference in outcomes.

We first investigate attrition rates by studying the percentage of the initial PIs who remained active in the NIH system and find that the attrition rate of the two groups differed significantly (Fig. 3a). In the year immediately following treatment, the near-miss group had 11.2% fewer active PIs than the near-win group (*p*-value < 0.001). This difference is not simply because near winners received an initial grant. Indeed, the gap persisted and extended beyond the first five years, remaining at 11.8% in year seven (*p*-value 0.002), followed by a drop afterwards. The RD analysis indicates that an early-career near miss on average led to a 12.6% chance of disappearing permanently from the NIH system over the next ten years (see Methods section). These results thus highlight the fragility of a junior scientific career, with one early near miss being associated with significantly higher attrition from the NIH system, despite the fact that to become an NIH PI, one had to go through years of training with a demonstrated track record of research. Notwithstanding the evidence that PhDs who left science are disproportionally employed at large, high-wage establishments[58], Fig. 3a documents differential survivorship between near winners and near misses, which raises the important next question: Could screening alone account for the observed performance advantage?

To understand the nature of the potential screening effect, we first test its underlying assumption by comparing pre-treatment characteristics of near-miss and near-win PIs who remained, finding a lack of difference between these two groups in any observable dimension *ex ante* (Fig. S22a), which suggests the screening effect, if any, may be modest (SI Appendix S3.7). To further examine potential screening effects, we next removed PIs from the near-win group, such that the attrition rate following removal is the same between the two groups (Fig. 3b). We performed a conservative estimation by removing PIs from the near-win set who, *ex post*, published the fewest hit papers but had the most publications. In other words, we created a subpopulation of near winners that had the same attrition rate as the near misses but are aided by an artificial upward adjustment to their hit probabilities (SI Appendix S3.7). We find that, while the performance of near winners improves by construction following this conservative removal procedure, the improvement is not sufficient to account for the observed performance gap. Indeed, in terms of the probability of producing a hit paper, or the average citations per paper, the near-miss group still outperformed the near-win group (Fig. 3cd, *p*-value < 0.001). The RD estimation yields consistent conclusions (Fig. S7d-f).

Together, these results demonstrate that the screening effect may have played a role, but it appears insufficient to entirely account for the observed difference between near misses and near winners.

We clarify these results further on several dimensions. To understand if the average improvement of the near misses masks heterogeneous responses, we measured the coefficient of variation for citations, finding a lack of difference between the two groups, suggesting a homogenous improvement within the group (SI Appendix S3.8, Fig. S16). We also compared the median citations to eliminate the role of outliers, yielding the same conclusion (SI Appendix S3.8). To rule out collaboration effects, whereby early-career setbacks might lead junior scientists to seek out advantageous collaborations, we restricted our analyses to lead-author publications only, and controlled for the status of their collaborators, yielding the same conclusions (Fig. S13, Fig. S20). Further, to check that the uncovered performance gap is not simply because near winners became worse, we selected a group of clear winners whose scores were further removed from the funding threshold. We find, as expected, that this group of PIs performed substantially better than the near-miss group prior to treatment. Yet, in the ten years afterwards, they show a similar performance as the near-miss group (SI Appendix S3.10), indicating that near misses performed at a comparable level as the group that appeared demonstrably better than them initially. We also repeated all our analyses by varying our definition of active PIs by focusing on publishing scientists only (SI Appendix S3.9, Fig. S18), and the definition of pay lines by using the NIH percentile score instead of priority score (SI Appendix S3.11). Amid all variations, our findings remain the same.

**Discussion**

Overall, these results document that an early-career setback has powerful, opposing effects, hurting some careers, but also, quite surprisingly, strengthening outcomes for others. As such, these findings show that prior setback can indeed be a mark of future success. Screening effects may partly be responsible yet appeared insufficient to explain the magnitude of the observed effects, supporting the idea that failure may teach valuable lessons[18-20,23].

The classic idea that "what doesn't kill me makes me stronger" may operate according to multidimensional mechanisms. We explored ten different observable dimensions, including shifts in intellectual direction and leadership, institutional locus, and collaboration patterns (see Methods section). We find only a suggestive tendency for near-miss scientists to publish in "hot topic" areas following treatment, although accounting for this factor did not reduce the observed performance gap (Fig. S21). More generally, these numerous observable features considered do not account alone or collectively for the performance change, suggesting that unobservable dimensions, including effort or grit factors following setbacks[20,22], may play a role. Crucially, the empirical findings and conclusions reported in the paper hold the same, regardless of the underlying processes. Indeed, while most empirical and theoretical evidence in science thus far documents that individuals benefit tremendously from success[1,2,4-8,27,50,55], our results offer among the first empirical evidence showing that some individuals can also benefit from setbacks, which may have broad implications for both individual investigators and institutions that aim to support and nurture them.

The design of our study necessitates the focus on near-miss individuals among all others who had setbacks[59,60]. While the RD approach allows us to expand our sample to a wider range of setbacks by controlling for evaluation scores, yielding the same conclusions (see Methods section), to what degree our findings may generalize substantially beyond near misses is a question we cannot yet answer conclusively. Moreover, the opposing effects of early setbacks also suggest there may exist population heterogeneities in responses that are worth exploring further. Who tends to be the most vulnerable, and who the most resilient? Quantitative answers to these questions may be crucial for the interpretation of our insights to inform policies and intervention strategies for building a robust scientific workforce[61]. It is also important to note that, while citations and their variants have been used extensively to quantify career outcomes[7,40,45,62-65], they represent an imperfect proxy at best to measure scientific outputs, suggesting that future research may fruitfully explore measurements that extend beyond citation-based measures to examine broader career outcomes.

Moreover, our analyses estimate the net advantage of near misses over near winners, which is only detectable if the gross advantage of early-career failure outweighs any benefits conferred by success. Given the widespread, convincing evidence supporting the validity of the Matthew effect in science[1,3-9,27,34] and beyond[3,10-13], where past success begets future success, these results suggest that powerful, offsetting mechanisms may be at work. This implies that, in areas where the Matthew effect operates less, the net advantage of failure may be more pronounced, suggesting that other domains provide important additional avenues for future work.

Finally, note that our results do not imply that one should strategically put roadblocks in the way of junior scientists, as the precondition for becoming stronger is to "not be killed" in the first place. The findings do suggest, however, that for those who persevere, early failure should not be taken as a negative signal—but rather the opposite, in line with Shinya Yamanaka's advice to young scientists, after winning the Nobel prize for the discovery of iPS cells, "I can see any failure as a chance."

## Methods

**Testing various generative processes.** While the broad idea of a setback-driven boost may take many forms, several such mechanisms may be detectable from data in our context. For example, (*A*) did early-career setbacks propel persistent junior scientists to attempt more novel research, whereas near winners were bound to their original ideas? (*B*) Did early-career setbacks lead junior scientists to seek out advantageous collaborations? Indeed, as teams are increasingly responsible for producing high-impact work[31,66], the observed performance gap might reflect collaborations, whereby near misses more frequently teamed up with higher-impact scientists and/or published fewer lead-author publications than their near-win counterparts. Alternatively, the uncovered difference might reflect an observable personal change process in terms of intellectual or physical mobility, as captured by research direction shifts (hypothesis *C*)[67] or changing institutions (hypothesis *D*)[68,69]. We tested hypotheses *A–D* from our data, finding that there is only a suggestive tendency for near-miss scientists to publish in "hot topic" areas following treatment, although accounting for this finding does not reduce the observed performance gap (Fig. S20, SI Appendix S4.5). However, none of the hypotheses alone can fully explain the observed performance gap between near misses and near winners (SI Appendix S4.1 – S4.4, Fig. S20); nor do these hypotheses combine to explain the findings when we control for all the factors outlined in hypotheses *A–D* together (SI Appendix S4.5). Overall, investigating these many dimensions narrows the potential interpretations of our results while further suggesting that additional sub-processes may be at work, including effort or grit factors following setbacks[20,22], which are difficult to observe directly from data and provide areas for future research. Crucially, the empirical findings reported in the paper hold the same, net of potentially many underlying processes.

**Econometric model specification and estimation procedures.** A possible concern with the comparison between the near misses and near winners near the threshold is endogeneity[50]; i.e., there might be other factors that influence both the funding decision and future career outcome. The finding that observable features of the two groups are statistically indistinguishable prior to the funding decisions helps diminish this concern. Further, were there some unobserved factor determining outcomes, such a factor would need to both advantage the near winners in getting initial funding yet disadvantage them over the longer run, which may be implausible. Nevertheless, to fully eliminate such endogeneity concerns, one needs to employ a causal inference framework called fuzzy regression discontinuity design (RD)[56]. In this section, we describe in detail our econometric model specification and estimation procedures.

The key idea of RD is that if decision rules create a jump in the likelihood of treatment, often at an arbitrary threshold, we can exploit this local discontinuity as an exogenous variable to predict outcome, instead of using the specific, realized treatment outcome, which could be influenced by endogenous factors. Such RD approaches have been widely used in the fields of education, labor, political economy, health, crime, and environmental studies[57], in addition to prior research on funding data[8,50,54,70].

In our setting, the likelihood of treatment (i.e., receiving funding) is largely determined by the score of each proposal[50,70]. The highly nonlinear relationship between the evaluation score and funding success (Fig. 1a) makes our fuzzy regression discontinuity design feasible. More specifically, we can treat whether scores fell just above or below the cutoff as an instrument to predict funding, and then use the *predicted* funding outcome, rather than the actual funding outcome, to predict future career outcome. The logic of this procedure is the following. Both the actual funding outcome and future career outcome can be affected by observable or unobservable factors. But since whether or not the score of the proposal is above or below the funding threshold is not influenced by any endogenous factors, if that as an instrumental variable can predict future career outcome, then it means there must exist a link from funding outcome to future career outcome, because the only way for the instrumental variable to influence future career outcome is through the funding outcome. Detailed causality diagram is shown in Fig. S6. Another advantage of using the instrumental variable approach is to allow us to control for the score itself (the so-called "running variable" in RD), and thus control for any distinctions in applications that vary smoothly with the score.

More specifically, we estimate the causal effect of early-career setbacks on future career outcomes using two-stage least squares regression (2SLS): in the first stage, the instrumental variable (being above or below the score threshold) is used to predict the funding outcome; in the second stage future career outcomes are regressed on the predicted funding outcome from the first stage. As illustrated in Fig. S6, the fuzzy RD approach eliminates any potential unobservable factors (e.g., novelty bias, hot topic, grit personality, perseverance, and so on) that might affect both the funding and career outcomes. This means, any significant results obtained from this approach can be interpreted as causal relationships.

More specifically, given the jump in the probability of receiving the funding at $s_0$ (i.e., normalized score 0), we have

$$P(F_j = 1|s_j) = \begin{cases} g_1(s_j) & if\ s_j \geq s_0 \\ g_2(s_j) & if\ s_j < s_0 \end{cases},$$

where $s_j$ is the running variable, i.e., normalized score for proposal $j$, $F_j$ is the funding decision outcome, and $g_1(s_j) \neq g_2(s_j)$ at $s_0$. The probability to receive treatment is

$$E[F_j|s_j] = P(F_j = 1|s_j) = g_2(s_j) + [g_1(s_j) - g_2(s_j)]z_j,$$

where $z_j = 1$, if $s_j \geq s_0$ and $z_j = 0$ otherwise. Since $s_0$ is the normalized score with a funding threshold that divides the proposals at an arbitrary point, the dummy variable $z_j$ is uncorrelated with any observed or unobserved factors[50]. To this end, we treat $z_j$ as the instrument variable and employ a simple two-stage least square (2SLS) regression estimation. Let $y_{it}$ be some scientific outcome of individual $i$ during time period $t$, i.e., number of publications, number of hit papers, or probability to publish hit papers. We conduct the estimation as follows:

*1st stage:* $F_j = \alpha_0 + \alpha_1 s_j + \alpha_2 s_j^2 + \cdots + \alpha_p s_j^p + \pi_i z_j + \theta X_{i,pre} + \mu_t + \eta_n + \eta_j,$

*2nd stage:* $y_{it} = \beta_0 + \beta_1 s_j + \beta_2 s_j^2 + \cdots + \beta_p s_j^p + \lambda \hat{F}_j + \gamma X_{i,pre} + \mu'_t + \eta'_n + \epsilon_i,$

where $X_{i,pre}$ is the prior performance of researcher $i$, i.e., prior number of publications, prior number of hit papers; $\mu_t$ and $\mu'_t$ are time fixed effects for application year, $\eta_n$ and $\eta'_n$ are NIH institution fixed effects, $p$ is the order of the polynomial control of the running variable, and $\eta_j$ and $\epsilon_i$ are error terms from the first and second stage, respectively. Moreover, $\hat{F}_j$ is the predicted values from the first stage, which are uncorrelated with the error term $\epsilon_i$, and $\lambda$ is the causal effect of near miss on future career outcomes. For a large sample that span a significant fraction of PIs around 0, we need careful controls of priority score. In the following analyses, we added the linear control of score when considering $\pm 10$ discontinuity sample ($p = 1$), and 3rd order polynomial control for $\pm 25$ discontinuity sample ($p = 3$). Finally, we eliminate these polynomial controls as we restrict the sample to the very narrow region around the discontinuity point[56] (in our setting, $\pm 5$ discontinuity sample).

The credibility of these estimates hinges on the assumption of the lack of prior knowledge of the cutoff, $s_0$, so that individual scientists cannot precisely manipulate the score to be above or below the threshold. This assumption is valid in our setting, because the scores are given by external reviewers, and cannot be determined precisely by the applicants. To offer quantitative support for the validity of our approach, we run the McCrary test[71] to check if there is any density discontinuity of the running variable near the cutoff, and find that the running variable does not show significant density discontinuity at the cutoff (bias=-0.11, and the standard error = 0.076). Together, these results validate the key assumptions of the fuzzy RD approach.

To understand the effect of an early-career near miss using this approach, we first calculate the effect of near misses for active PIs. Using the sample whose scores fell within -5 and 5 points of the funding threshold, we find that a single near miss increased the probability to publish a hit paper by 6.1% in the next 10 years (Fig. S7a), which is statistically significant ($p$-value < 0.05). The average citations gained by the near-miss group is 9.67 more than the near-win group (Fig. S7b, $p$-value < 0.05). By focusing on the number of hit papers in the next 10 years after treatment, we again find significant difference: near-miss applicants publish 3.6 more hit papers compared with near-win applicants (Fig. S7c, $p$-value 0.098). All these results are consistent with when we expand the sample size to incorporate wider score bands and control for the running variable (Fig. S7a-c).

For our test of the screening mechanism, we employ a conservative removal method as described in the main text and redo the entire regression analysis. We recover again a significant effect of early-career setback on the probability to publish hit papers and average citations (Fig. S7d-e). For hits per capita, we find the effect of the same direction, and the insignificant differences are likely due to a reduced sample size, offering suggestive evidence for the effect (Fig. S7f).

**Acknowledgements** The authors thank A.-L. Barabási, J. Chown, J. Evans, E. Finkel, V. Medvec, J. Loscalzo, W. Ocasio, P. Stephan, B. Uzzi, Y. Yin, and all members of Northwestern Institute on Complex Systems (NICO) for invaluable comments. This work is supported by the Air Force Office of Scientific Research under award number FA9550-15-1-0162 and FA9550-17-1-0089, Northwestern University's Data Science Initiative, the National Science

Foundation grant SBE 1829344, and Alfred P. Sloan Foundation Award G-2015-14014. This work does not reflect the position of NIH.

**Correspondence** Correspondence and requests for materials should be addressed to D.W. (email: dashun.wang@northwestern.edu).

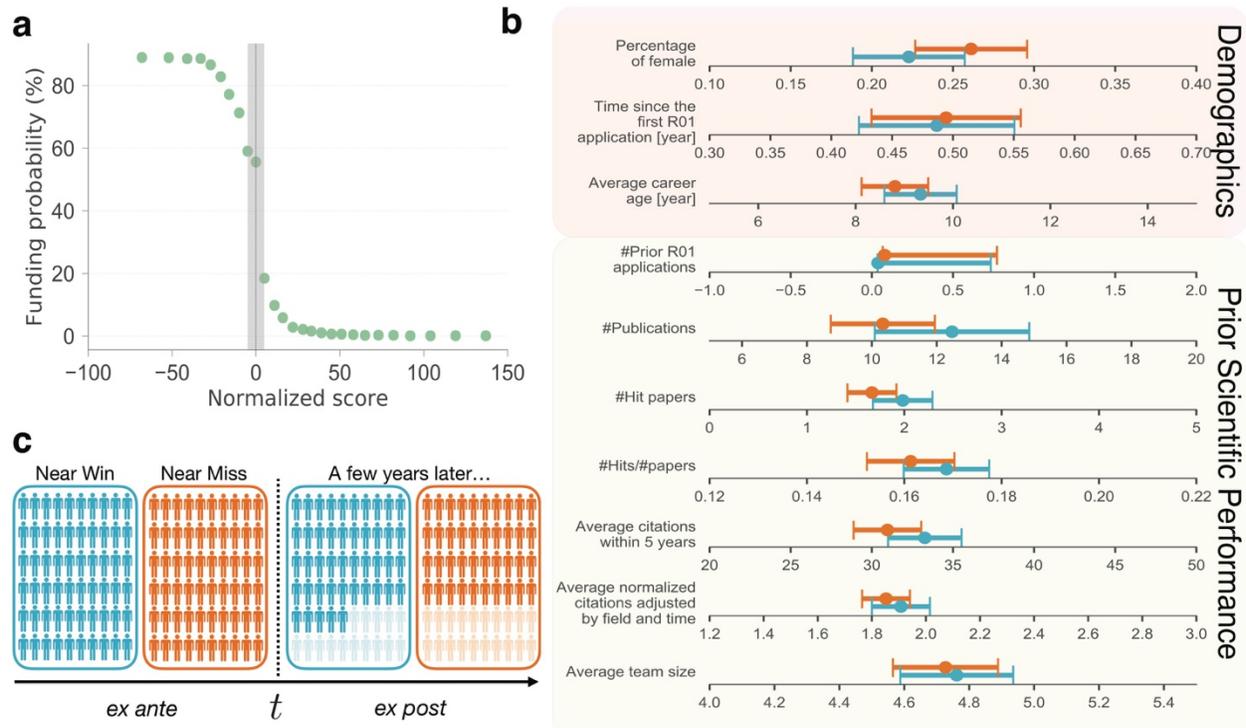

**Figure 1: Pre-treatment comparisons between the near-win and near-miss applicants.** (**a**) Relationship between normalized score and award status. Funding probability shows a clear transition around the funding threshold. We focus only on junior PIs whose normalized scores lie within the range (-5, 5), the shaded grey area, which includes 561 near-win and 623 near-miss applicants in our sample. (**b**) Pre-treatment feature comparisons between the near-miss and near-win group. We compared 10 different demographic and performance characteristics. The features are defined as follows (from top to bottom): 1) percentage of female applicants; 2) number of years since the first R01 application; 3) number of years since the first publication; 4) number of previous R01 applications; 5) number of publications prior to treatment; 6) number of prior papers that landed within the top 5% of citations within the same field and year; 7) probability of publishing a hit paper; 8) average citations papers received within 5 years of publication; 9) citations normalized by field and time[32]; and 10) average team size across prior papers. We see no significant difference between the two groups across any of the ten dimensions we measured; Error bar represents the 95% confidence interval. (**c**) An illustrative example of the underlying process. Solid color indicates people who remained active, whereas shaded color denotes the fraction that disappeared from the NIH system. Blue and orange indicate near-win and near-miss applicants, respectively.

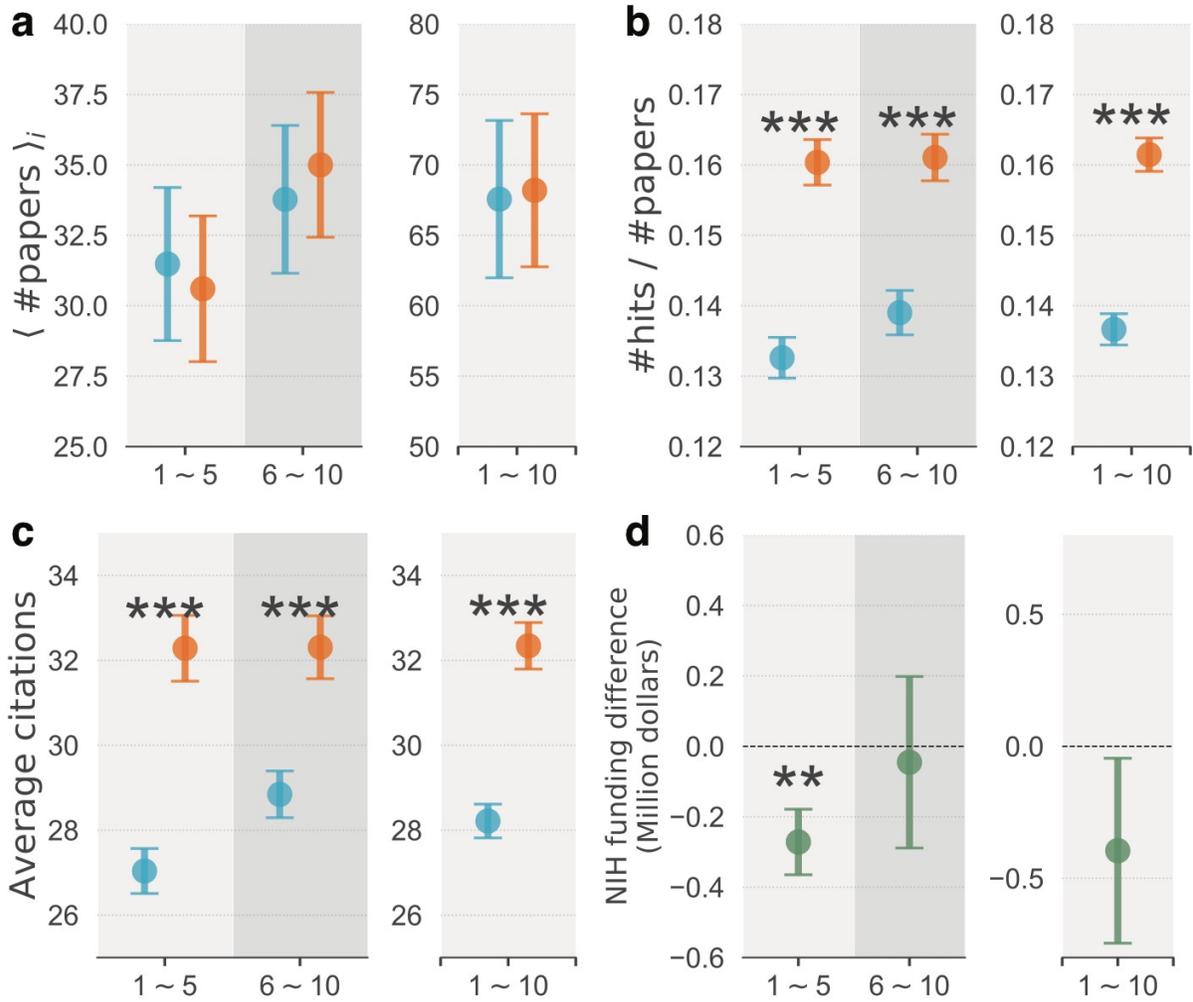

Figure 2: **Comparing future career outcome between near misses (orange) and near winners (blue).** (a) The average number of publications per person. (b) Near misses outperformed near winners in terms of the probability of producing hit papers in the next 1-5 years, 6-10 years, and 1-10 years. Note that there appears a slight performance improvement for the near-win group in the second five-year period, but the difference is not statistically significant ($p$-value > 0.1). (c) Average citations within 5 years of publication. The near-miss applicants again outperformed their near-winning counterparts. To ensure all papers have at least 5 years to collect citations, here we used data from 1990 to 2000 to avoid any boundary effect. (d) Funding difference between the near-miss and near-win group from the NIH (near misses minus near winners). *** $p < .001$, ** $p < .05$, * $p < .1$; Error bars represent the standard error of the mean.

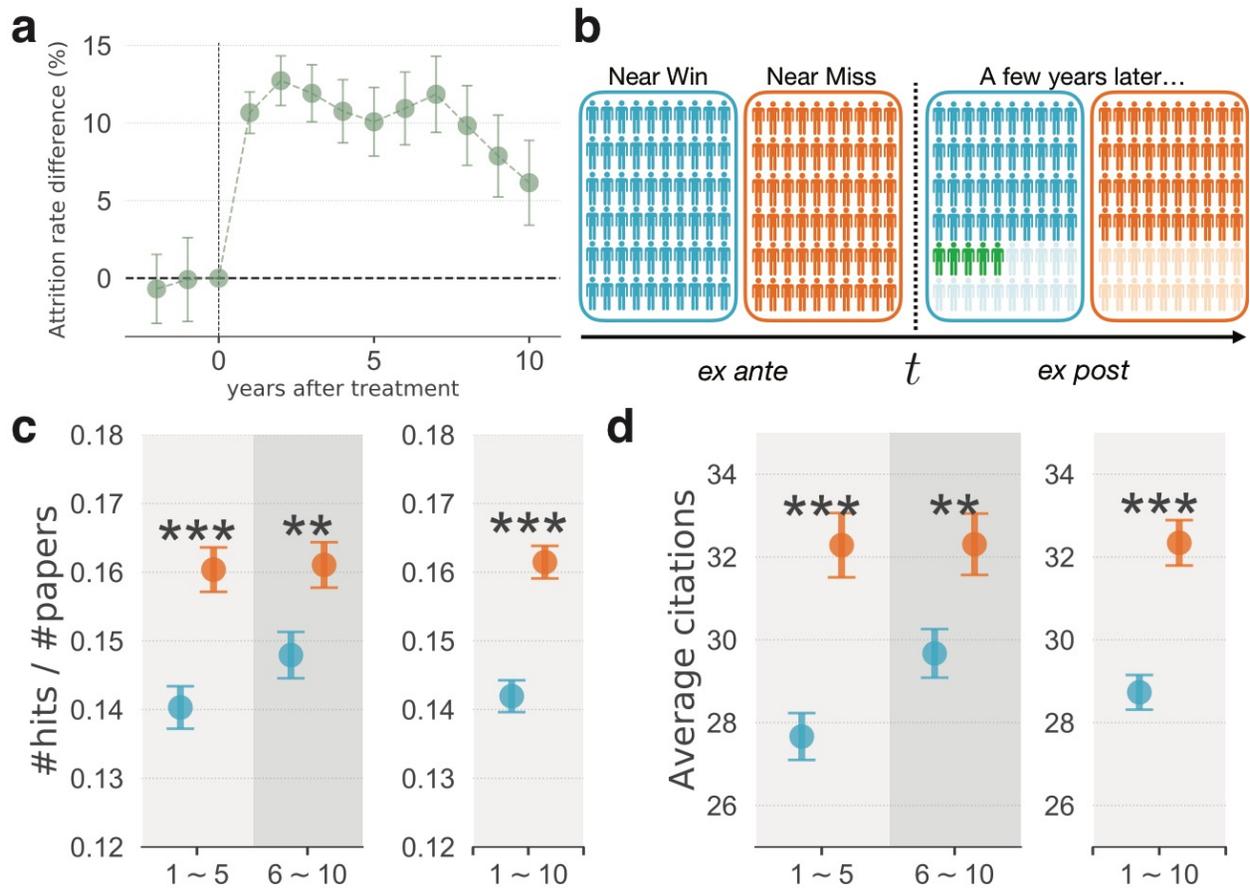

**Figure 3: Testing the screening hypothesis with a conservative removal procedure.** (**a**) Attrition rate difference between the near-miss and near-win group (near misses minus near winners). We measure the percentage of PIs remained in each of the two groups, and calculate their difference in each of the ten years after treatment. (**b**) An illustration of the conservative removal procedure. To test if the observed performance difference can be accounted for by the population difference, we performed a conservative estimation by removing PIs who published the fewest hit papers but with the most publications from the near-win group (blue), such that after removal (green) the two groups have the same fractions of PIs remaining. After removal, the near-miss group still outperformed the near-win group in terms of the probability of producing a hit paper (**c**), or the average citations of papers (**d**). The results shown in **c-d** demonstrate that while the performance of near winners indeed improved following the conservative removal procedure, the screening hypothesis alone cannot account for the uncovered performance gap. *** p < .001, ** p < .05, * p < .1; Error bars represent the standard error of the mean.

# Supporting Information

# Early-career setback and future career impact

Yang Wang [1,2], Benjamin F. Jones [1,2,3], and Dashun Wang [1,2,4] *

[1] *Northwestern Institute on Complex Systems, Northwestern University, Evanston, IL, USA*
[2] *Kellogg School of Management, Northwestern University, Evanston, IL, USA*
[3] *National Bureau of Economic Research (NBER), Cambridge, MA, USA*
[4] *McCormick School of Engineering, Northwestern University, Evanston, IL, USA*

* To whom correspondence should be addressed. e-mail: dashun.wang@northwestern.edu

Table of Contents

# S1 Data Description

In this paper, we assembled large-scale datasets from a range of different sources: 1) the U.S. National Institutes of Health (NIH) grant application data; 2) Thomson Reuters' Web of Science data (WoS); 3) the National Library of Medicine's PubMed database; and 4) Google Scholar (GS) profiles. Combining these datasets allows us to trace career trajectories of NIH principal investigators (PIs) through their grant applications, funding outcomes, scientific publications and publication impacts.

## S1.1 NIH grant database and sample details

Our main dataset contains all R01 grant applications ever submitted to the NIH between FY1985 and FY2015; R01 is the primary funding mechanism in the world's largest public funder for biomedical research. Our data consist of 778,219 competing grant applications in total, supporting more than 170,000 research personnel across more than 2,500 U.S. institutions. For each grant application, we obtained its evaluation score, scientific study section, a unique identifier for the PI, the PI's name, and the funding outcome. Figure S1 shows the number of competing R01 applications and the success rate over time; the success rate has declined over the past three decades, due to the increased number of applications with limited budgets to support them (Fig. S1b).

## S1.2 Publication and citation datasets

Our second dataset, i.e., the Web of Science, provides comprehensive publication and citation records of more than 46 million papers published from 1900 to 2015 by more than 20,000 different journals. For each paper, the WoS data contain the title, journal, subject, publication date, author names, affiliations, and a set of references.

For every NIH grant, all papers published as a result of the grant are deposited in the PubMed database and uniquely identified with PubMed IDs. This fact enables two important possibilities. First, it allows us to link the NIH grant database with the WoS dataset using a precise mapping between PubMed IDs and paper identifiers in our WoS database purchased through an advanced agreement with Thomson Reuters. This mapping not only offers us additional information about each publication, it also allows us to trace citations to each paper within the WoS and how these citations compare with other papers published in the same year and field. Second, each PI is associated with his/her unique IDs in the NIH data, and the PubMed papers associated with all his/her grants offer a set of "ground-truth" papers for disambiguation purposes. As such this key feature of our data could also substantially improve the accuracy with which we identify authors in the publication databases, as we will discuss in detail in S1.4.

## S1.3 NIH grant review process

To make sure all readers are familiar with the process through which our study took place, here we briefly review the NIH grant review process[1]: In the first step, each PI submits his/her grant proposal directly to the Center for Scientific Review (CSR), and a referral officer will assign the application to an appropriate Integrated Review Group (IRG), Scientific Review Group (SRG) and Institute or Center (I/C). After assignment, each application receives an identification number. All applications will be assigned to a review group (known as a study section) that comprises scientific peers, where reviewers usually evaluate the proposal according to several metrics such as innovativeness, significance, investigator's track record, et cetera. About 50% of all applications are rejected prior to the panel discussion step, and thus no score is assigned. For applications that reach the discussion panel, each study section member provides a priority score from 1.0 (best) to 5.0 (worst) with increments of 0.1, and the score is averaged and multiplied by a factor of 100. These scores are then normalized within each study section to facilitate funding decisions.

In this study, we focus on all new competing grant applications from regular standing study sections between FY1990 and FY2005. We did not use data from FY1985 to FY1989 to eliminate possible boundary effects in defining junior PIs. Similarly, we focus on grants prior to FY2006 because we need additional ten-year time window to trace career outcomes. To further make sure that near-miss and near-win individuals studied in our paper apply to categorically similar R01 programs, we eliminated special-emphasis panels (special emphasis panels are review groups formed on an ad hoc basis to review applications requiring special expertise or when a conflict of interest situation occurs), as well as small study sections with less than 50 grant applications per year, to ensure statistical power when making inferences about proposals close to the funding score threshold.

## S1.4 Author name disambiguation by combining WoS and NIH datasets

Each PI in the NIH is uniquely identified by an NIH ID together with his or her full name, and all publications resulting from his/her NIH grants are correctly associated with the PI. This feature provides useful information that helps with author name disambiguation. For any paper in a scholarly database, a fundamental issue is to identify the individual(s) who wrote it and, conversely, to identify all of the works that belong to a given individual[2-4]. This seemingly simple task represents a major unsolved problem for information and computer sciences, mainly because of the lack of ground truth. While automatic name disambiguation in large-scale scholarly datasets remains an unsolved challenge, recent studies on individual careers have achieved initial success using various pertinent citation features[2,5]. In this paper, we build on the state-of-art algorithms but also extends them in meaningful ways by combining the NIH dataset with the WoS dataset. Indeed, the NIH publications associated with each PI represent new, ground-truth information that none of the existing disambiguation algorithms had access to. When working in unison with existing algorithms, such information allows us to further harvest additional publications by the PI, which could potentially offer a more accurate and comprehensive trace for individual career histories than simply using the state-of-art algorithms. Next we describe specific steps in our name disambiguation procedure.

First, for each PI name we generated a pool of candidate publications, consisting of all possible publications by the PI from the WoS according to the following rules: a) Last names of these publications must be the same with the PI last name; b) Initials of the first names have to be the same with the first name of this PI. If full first names from the WoS are available, they must be identical to this PI's first name; c) The same rule applies to middle names. One can think of the goal of this step as to increase the rate of recall at the expense of precision.

Our second step involves merging papers from the pool generated above. Here we build on existing algorithms[2,6,7] to decide the authorship of papers: Two papers are considered to be from the same author if one of the following conditions is true: a) The two papers cite each other; b) The two papers share at least one common reference.

After step two, we compare papers from each group with NIH supported publications by the considered PI, following these two rules: a) If the PI has at least one NIH-related publications, all groups that contain these publications are associated with this PI; b) If there is no NIH-related publication for a particular PI, we compared the string similarity between PI's affiliation on the NIH grant applications and on the publications, after removing common words (e.g., stop words, "university", "institute", "hospital", etc.), and consider the paper belongs to the PI for the cases with high cosine similarity (in our case, we set the threshold at 0.65), as well as its group.

Although name disambiguation methods are becoming increasingly accurate, the main problem with them is that we are often not sure how perfect or imperfect they are. To gauge the accuracy of our algorithm, we compare career histories obtained using our method with Google Scholar (GS) individual profiles. GS allows individual scientists to create, maintain, and update their own profile of publication records, assisted by the name disambiguation algorithm developed by Google. While the GS profiles themselves are not gold standards, here we quantify the extent of agreement between our method and GS to make sure that our name disambiguation method does not deviate significantly from an independent test. Comparing with GS profiles, we find that the average precision and recall of our algorithm is 0.85, and 0.71, respectively. An alternative way that we could make use of the GS profiles is to incorporate them directly into our disambiguation procedure. The reason we did not do that is to avoid potential bias in our results, because not everyone in our sample has a GS profile. Indeed, we find that only a fraction of NIH PIs has reliable GS profiles, which are biased towards active PIs. Since the goal of our study involves testing differences between two populations, the potential bias introduced by the GS profiles could be an issue as more near winners remained active in our setting.

While the analyses in the preceding paragraph show a high degree of agreement between our method and an independent test set, as with any name disambiguation methods, it is necessarily imperfect. To ensure the errors contained in the method do not affect the conclusions of our study, and to examine further the robustness of our results, we repeated our analyses by varying different assumptions made in the disambiguation procedure (Fig. S2). For each variation, we redo the disambiguation process and repeat our analysis to see if the conclusions change. For example, in Method A, we considered only self-citations between two papers to determine if they are from the same author; In Method B, we used not only self-citations and co-references, but also take into account the names of coauthors. For example, if two papers associated to the same name share at least one common coauthor name, or one common reference, or cite each other, they are considered to be from the same PI. These different methods result in different set of papers for each author. Yet we find, amid all different variations, our conclusions remain the same (Fig. S2). This supports the hypothesis that while name disambiguation methods may be error-prone, the errors do not affect the conclusion if they are distributed unbiased across the two populations. Indeed, we find that, under different disambiguation methods, near misses consistently have a higher chance to publish hit papers (Fig. S2a and e). Near misses also outperform near winners in terms of the average number of citations per paper (Fig. S2b and f). Moreover, near misses publish significantly more hit papers per person than near winners using Method A (Fig. S2c). In terms of the number of publications, as in the main text there is again a lack of significant difference between the two groups (Fig. S2d and h, Fig. 2a in the main text). Finally, PIs with common Chinese or Korean names are known to be difficult to disambiguate, accounting for the majority of the error rate[2,3,8]. As another robustness check, we repeated our analyses by removing PIs with common Chinese or Korean last names (Fig. S3), finding the results are similar to those reported in the main text. While the name disambiguation methods may be imperfect, the goal of this paper is to incorporate a range of pertinent features and to try many different variations. The consistency across all these analyses helps document the robustness of our results.

## S1.5 Normalized Priority Score

In this study, we take advantage of the fact that NIH funding decisions are based on evaluation scores that are observable from data, resulting in a highly non-linear relationship with funding success. The funding cutoff at the NIH varies across different study sections, institutes, year and grant mechanisms. To infer the cutoff, here we build on prior work[9] to obtain the priority score for each application. In addition, our data allowed us to zoom into each study section instead of focusing on the NIH institute level. This helps us account for the heterogeneity of cutoff across different years and study sections, by comparing, for each fiscal year, the grant applications submitted to the same study section. We investigated the success probability as a function of priority score and define the cutoff score as the third worst score of funded applications in each study section so that there are only two out-of-order funding events that occurred beyond the cutoff. This choice helps balance the number of observations in both near-win and near-miss group. To check if it indeed leads to a cutoff effect, we calculate the normalized score by subtracting the cutoff score from each proposal score, finding a clear drop in funding success probability at the cutoff point (Fig. S4 and Fig. 1a); the same results hold for major NIH institutes (Fig. S4b). We also repeated our analyses by defining the cutoff as the second worst score of funded applications in each study section, finding consistent conclusions (Fig. S5).

## S2 Related Work

There has been extant theoretical and empirical evidence from a diverse set of disciplines that can help inform our research question of whether an early-career setback may lead to future career impact. Here we aim to provide a brief overview of the different lines of research.

An especially prominent and relevant line of enquiry is the Matthew effect, which was formalized by Robert K. Merton in 1968[10], named after a verse in the bible's Gospel of Matthew. Merton described a "rich get richer" phenomenon, where early-career success brings reputation and recognition that can translate into tangible assets that in turn help bring future successes. The idea of the Matthew effect has deep roots, and goes under many different titles, including cumulative advantage[11] used by de Solla Price to explain citations of papers, and preferential attachment[12] in the network science context to explain the origin of scale-free networks. The idea also forms the core of various fundamental, mathematical models, including Pólya process, corresponding to the well-known urn model by György Pólya in 1923, or the Yule process[13]. It also underlies the key assumption used by Zipf to explain the fat tailed distribution of wealth in the society[14].

The Matthew effect predicts that the answer to the question of whether early setbacks may lead to longer-term success should be no. This thesis is further supported by widespread empirical evidence in science as well as in other related domains. For example, for publications, the Matthew effect captures the well-documented fact that highly-cited papers are more visible and are more likely to be cited again in the future[11,15-19]. Studies of individual scientific careers documented the reputation effect that an early-career recognition brings to increase future chances of success[20-23]. Beyond science, the Matthew effect has also been experimentally validated in various domains, from social influence[24] to different reward systems[25,26]. It also speaks to a classical line of research in the literature of motivation[27], whereby success bolsters confidence whereas failure lowers self-esteem.

The strongest endorsement to the Matthew effect comes from the fact that it has been investigated specifically in the setting of our study over the past decade[9,28,29]. For example, Jacob and Lefgren first documented the Matthew effect in the science funding system[9]. Using data from the NIH, they found that a small but significant impact of initial NIH funding on obtaining future funding from the NIH, with little impact on future scientific productivity (i.e., number of publications). Ganguli investigated the impact of grants using data from a Soviet grant program shortly after the end of the USSR[28]. Her analysis directly speaks to the scenario when funding levels are low, and found that obtaining grants significantly reduces the attrition rate while increases future productivity. The most recent evidence of the Matthew effect in science funding was offered in the context of early-career academic funding system in the Netherlands[29]. More specifically, Bol et al. demonstrated that early career funding is critical for late-career funding, as winners just above the funding threshold accumulated more than twice as much as funding in the following eight years. Although in the Dutch setting, prior success of getting funded is a merit review criterion in later competitions which by itself may increase future chance for funding, these results are consistent with another recent study on the effect of postdoctoral fellowships[30]. where researchers showed that securing a specific postdoctoral fellowship from the NIH increases subsequent chances of getting another NIH grant. Together, the consistent results across all these studies offer clear, convincing endorsement that establishes the Matthew effect as the leading hypothesis for the research question raised in our paper. These studies are what make our findings surprising, thus highlighting the main novelty of our paper.

While the Matthew effect predicts that success breeds success, there are reasons to believe that the opposite may be true. The primary mechanism for this second school of thought is the screening effect, which builds on the observation that later competitions are held only among those who managed to stay in. Hence if early career setbacks increase attrition, it may act as a screening mechanism, leading to a more selected set of individuals who remain[31,32], with fixed, advantageous characteristics such as commitment, grit, high self-confidence, etc. As such, past failure may become a marker for future success for those who managed to stay in. This screening mechanism may produce empirically similar observations as the learning-from-failure literature, suggesting that failure may teach valuable lessons that is hard to learn otherwise[33-35]. In a separate line within the motivation literature, it has been shown that setbacks may also motivate people by signaling that more effort is needed, whereas success may offer a sense of partial goal attainment, signaling that less effort is needed to reach similar targets[36,37].

Finally, it is important to note that, while these different schools of thought make opposite predictions, they are not exclusive to each other. Rather they may both operate at the same time. Therefore, the net effect, which is what we can observe from data, remains unclear; this is the main purpose of our study.

## S3 Further Robustness Checks and Additional Results

In this section, we reported further robustness checks and additional results by constructing different experimental settings and parameters.

### S3.1 Hits per capita

In this section, we compared near misses with near winners in terms of hits per capita, measuring the average number of hit papers per person. We find a similar size of performance increases for near misses (near misses outperformed near winners by a factor of 19.2%, i.e., around 2 hit papers per capita over the next ten-year time window). The lesser significance level is due to a reduced sample size (Fig. S8, *p*-value 0.107). Moreover, the RD analyses yields similar magnitude of results (a single near miss leads to 3.6 more hit papers per capita over the next ten-year time window, *p*-value 0.098, Fig. S7c).

### S3.2 Alternative definitions of junior PIs

In the main text, we defined junior PIs as those within the first three years of their R01 NIH applications. To ensure our results are not affected by this definition, here we tried two variations of alternative variations and repeat all our analyses. First, we modified our definition of junior PIs by only focusing on first-time applicants, i.e., those who submitted their first R01 application at the time of treatment. This results in a corpus of 656 near win and 703 near miss applicants. We conducted the same analyses, finding the results are robust across various measures of longer-term success (Fig. S9).

We also varied the junior PI definition by focusing on those without any existing NIH grants at the moment of treatment within our main sample. We recovered the same findings for this group of PIs. For those who managed to stay in, their subsequent publications again garnered higher impacts in terms of probability of producing a hit paper (Fig. S10a), average citations (Fig. S10b), or hit papers per capita (Fig. S10c). To test the screening effect, we repeated our conservative removal procedure, finding again that near misses significantly outperformed near winners (Fig. S10d-f), with a stronger effect than what we observed in Fig. 3 of the main text (shaded symbols in Fig. S10d-f). Together these results indicate that our results are robust against different definitions of junior PIs.

## S3.3 Varying thresholds for the definitions of hit papers

To test if our results only hold for publications that fall within the top 5% of citations, we varied the definition of hit papers using top 1%, top 10% and top 15% of citations received in the same field and year (Fig. S11). Consistent with our main results, both near winners and near misses have significantly higher hit paper probabilities relative to the base rate, independent of our definitions. In all three different definitions, near misses outperformed near winners, demonstrating that the hit rate per paper by the near misses are significantly higher than that of near winners. In terms of hits per capita, near misses published significantly more top 1% highly cited papers than near winners. Using our conservative removal method, we arrived at the same conclusion that the screening mechanism alone seems insufficient to explain the observed difference.

## S3.4 Normalized citations over time and disciplines

Publications in some disciplines may be cited more frequently than in other disciplines[38]. To normalize the raw citations with respect to different disciplines, we follow the canonical method[38] and normalize the citations of each article by the average citations of all articles belonging to the same field and year. Specifically, let the raw citation of our focus article be $c$, and the mean citations for all articles in the same field and year be $c_0$. Then the normalized citations can be calculated as $c_f = c/c_0$. Moreover, we define field according to the citation network of each article. By dividing the average citation rates by an expected citation rate from papers in the same field, we calculated the relative citation ratio (RCR) for each article[39].

To show our results are robust with different citation measurements, we compare the average $c_f$ and RCR of articles published by near winners and near misses (Fig. S12). We find that near misses again outperformed near winners in terms of the normalized citations and the RCR. Both groups have substantially higher citations compared with the average citations of all articles in the same field and same year. In the first five years after treatment, near misses attracted 15% more citations compared with near winners, and this difference persisted for the next five years. Consistent with our main finding, the screening effect seems insufficient to account for the entire difference (Fig. S12bd).

## S3.5 Robustness to alternative fiscal years

To rule out the possibility that temporal effects drive the differences between near winners and near misses, in this section we focus on two sub-periods to see if our results hold in both time windows. We focus on two different time spans, from 1990 to 1995 and from 1995 to 2000. We find the results are robust with respect to different time periods (Fig. S14): in terms of hit rate per paper and average citations, near misses significantly outperformed near winners; in terms of hits per capita, we obtain the same direction with lesser significance level. There is no significant difference in terms of the number of publications per person.

## S3.6 Additional funding from the National Science Foundation (NSF)

While NIH is the world's largest public funder for biomedical research, there may still exist the possibility that near-miss PIs found funding elsewhere. Since we do not have data access to all possible funding agencies, we cannot entirely rule out this hypothesis. But here we provide a test case by using all NSF funding data to compare whether there is any funding difference between near misses and near winners. Note that NIH and NSF in total support a vast majority of funding obtained by biomedical junior scientists[40], and support 70% U.S. publications from 2008 to 2015. Specifically, we compared the total NSF funding amounts of NSF PIs with the same name as near misses and near winners in the same time period (Fig. S15). We find no significant difference in funding from the NSF between the two groups in the next 10 years.

## S3.7 On the screening mechanism

In this section, we conduct further analyses to test the underlying assumption of the screening mechanism. The screening mechanism hypothesizes that near misses who remained active in the next 10 years after setbacks are "better" *ex ante* than near misses who have been screened out. To test this hypothesis, we compared several pretreatment features between active and inactive near misses, finding near misses who have left the system are indeed weaker than those who manage to stay in (Fig. S22b). However, we conducted the same analysis for near winners, obtaining the same results (Fig. S22c), suggesting screening effect may occur, but if so, it runs in the same direction for both near winners and near misses. Indeed, we compared pre-treatment characteristics of near misses and near winners who remained active for the next ten years, finding a lack of difference between these two groups in any observable dimension *ex ante* (Fig. S22a). We also obtained the same results using 2SLS regression analysis with dependent variable being the probability to publish hit papers and average citations per paper *ex ante* (for both cases, $p$-value > 0.25). Together these results suggest that the screening effect might be modest, if exists at all.

Nevertheless, we further conducted the conservative estimation by removing "less able" near-winners to create artificial upward adjustment for performance of those who remained. As shown in the main text, we find upon removal, the performance gap remained. In addition to the removal procedure outlined in the main text, we also tested other, more conservative methods by redefining the "less able" PIs based on the LHS against which we compare the two groups. That is, when comparing hit rate per paper, "less able" PIs are those publish the fewest hit papers with the most papers; when comparing the number of hit papers and the number of publications, "less able" PIs are those who publish the least hit papers or publications, respectively. Amid all these variations, we find the main result holds the same (Fig. S23).

## S3.8 Variance and outliers

Previous sections show that near misses show an average increase in citations, here we examine other measures to inform further the shift in distribution. First, near misses have lower chance to publish low quality research whose citation is below the average of all publications from the same field and same time (Fig. S16a, $p$-value < 0.001). Second, we examine the coefficient of variation for citations of papers published by near winners and near misses in the next 10 years. The coefficient of variation (CV) is defined as $c_v = \frac{\sigma}{\mu}$, where $\sigma$ is the standard deviation of the citation, and $\mu$ is the average citations. We see that near-miss applicants have a slightly higher coefficient of variation, but the difference is not significant (Modified signed-likelihood ratio test (SLRT) for equality of CVs $p$-value > 0.1, Fig. S16b). Finally, we further compared the median citations in order to eliminate the effect of outliers. Within the next ten years, the median citations of the near misses are substantially higher than that of near winners, and the difference between the two groups is statistically significant (two sample Mood's median test $p$-value < .001). This is also true for the case of the conservative removal procedure (two sample Mood's median test $p$-value 0.017).

## S3.9 Different definitions of active PIs

In the main text, active PIs are measured through individual grant activities. Specifically, we trace all NIH grant activities for each PI, and define active PIs as those who applied for and/or received NIH grants at some point in the future, i.e. after the measurement time period. Correspondingly, inactive PIs are those who neither applied for NIH grants nor had one beyond the measurement time period, identifying those who disappeared from the NIH system. Formally, let us denote the treatment time by $t$, and our measurement period by $T_m$ ($T_m \leq 10$ years). A PI is considered as an active scientist in the NIH system she/he either applies for an NIH grant or receives one after $t + T_m$; she/he is defined as inactive otherwise. Note that this definition is a cumulative measurement. That is, once an NIH PI became inactive, he or she was never active again within the observation window covered by our data.

Finally, as an alternative robustness check, we modified the definition of active PIs by focusing on those who published at least one paper during a certain time window, i.e., PIs with at least one publication between $t$ and $t + T_m$. We find the results are robust in terms of this alternative definition (Fig. S18). The near misses again outperformed the near winners in terms of hit rate per paper and average number of citations; the effect cannot alone be explained by the screening effect as the difference of these two groups is still significant after conducting the conservative removal.

## S3.10 Was it because near winners became worse?

In the main paper we documented a performance difference between the near-miss and near-win groups. But could it be simply because near winners became worse after the treatment? To test this hypothesis, we investigated another group of clear winners. We selected success applicants whose scores were further removed from the funding threshold (normalized priority score range from -20 to -10). We find that prior to treatment, across many metrics, there is a clear difference between the near misses and clear winners. For example, in the prior three years before treatment, clear-win applicants typically have a 1% higher hit rate per paper compared with near-miss applicants; the number of hit papers per person is 2.3 compared with 1.7 for near misses; the average citations for the clear winners is 13% higher than for near misses; and clear winners show more research experiences than near misses. This advantage is expected given our design, showing that the clear winners clearly outperformed the near misses prior to treatment. After the treatment, however, we find that near misses outperformed clear winners in terms of hit rate per paper if we only focus on active PIs. With the conservative removal, we find the two groups have very similar performance 10 years after treatment in terms of different metrics (hit rate per paper, number of hit papers per capita, number of publications and average normalized citations in the 10 years after treatment, Fig. S17), indicating that near misses are now performing at a comparable level as the group that were demonstrably better than them prior to treatment. Given the fact that near misses are comparable with clear winners, who showed significant advantages *ex ante*, the observed gap is unlikely solely due to the fact that near winners become weaker.

## S3.11 Comparing near winners and near misses using the percentile score

NIH uses percentile score of proposals in each NIH institute to determine the funding outcome. A percentile score of an application is calculated by comparing all applications by the same study section at its last three meetings[1], which could take place in different years. Prior studies [9] used the priority score instead of the percentile score to infer the cutoff, which is what we did in the main text. But, to ensure that the results still hold, here we repeated our analyses by using the percentile to define the cutoff in each NIH study section. Specifically, for each study section in each fiscal year, we define the cutoff score as the third worst percentile score of funded grant applications in each study section so that there are only two out-of-order funding applications beyond the cutoff. Because percentile score has a narrower range than prior score, we focus on applications around the cutoff (from -2 to 2). By comparing the near misses with near winners, we find near misses consistently show higher performance in the longer run (Fig. S19).

## S4 Potential generative processes

While the broad idea of a failure-driven boost may take many forms, several such mechanisms may be detectable from data in our context. There are several plausible hypotheses for such processes as we mentioned in the main text. Here, we demonstrate how we test these hypotheses.

### S4.1 Hypothesis *A*: Novelty

Did early-career setbacks prompt near misses to attempt more novel research, whereas near winners are bound to their original ideas? Existing studies have shown that high-novelty research tends to attract more citations in the long-run[41], especially when a small degree of highly novel combinations of prior knowledge are balanced with otherwise highly conventional combinations of prior knowledge[42]. To test this hypothesis, we analyzed: 1) whether near misses are more likely to publish high-novelty papers in the ten-year window following treatment compared to near winners; 2) whether near misses shift toward publishing high-novelty papers compared to their own publications prior to treatment, relative to any such shift among near winners. By calculating paper novelty and conventionality based on its combinations of prior work[42], we conducted a 2SLS regression with the dependent variable being the probability of publishing a high-novelty paper. In a separate regression, we study the tendency to produce high-conventionality papers in the next ten-year window. We find there is little evidence that the early-career setback may have any effect (Fig. S21, *p*-value = 0.291 for novelty; *p*-value = 0.51 for conventionality). Further, we tested if there is any effect of early-career setbacks on publishing articles that include the highly novel and highly conventional combinations of prior work that have been shown elsewhere to predict high impact, finding again insignificant results (*p*-value = 0.67).

Finally, we introduce a standard difference-in-difference specification, measuring the shift of near misses toward publishing novel papers, compared to what the near misses used to do and relative to near winners. More specifically, we run a regression with interaction terms to capture any differential shift among near misses:

$$P(high\ novelty\ publication) = \alpha\ near\_miss + \beta T_{post} + \gamma\ near\_miss * T_{post} + \epsilon,$$

where $near\_miss = 1$ if the PI is a near miss, 0 otherwise; $T_{post} = 1$ if after treatment, 0 otherwise. We find, comparing with near winners, near misses were no more likely to publish high-novelty papers relative to their prior publications (coeff. = 0.02, standard error = 0.017, *p*-value = 0.302). We obtained similar results when focusing on the likelihood of publishing highly conventional papers or balanced high-novelty/high-conventionality papers (*p*-value > 0.1 in both cases). Overall, the effect of the early-career setback does not seem attribute to the hypothesis that near misses published more novel papers.

## S4.2 Hypothesis *B*: Collaboration effect

Did early-career setbacks lead junior scientists to seek out advantageous collaborations? Previous studies have shown that teams are often responsible for producing high-impact work[43,44], suggesting that a collaboration effect may be a plausible explanation for the performance gap. Having missed out on funding, the near misses may resort to collaborations with other authors; hence the increase in hit papers we observed may potentially reflect work performed with distinguished colleagues. To test this hypothesis, we first repeated our analyses by focusing only on lead-author publications for the two groups. Here we follow existing studies[45] and define lead-author publications as the first-author or last-author publications. We find that, by focusing on lead-author papers only, our conclusions remain the same (Fig. S13a-d). We further employed the conservative removal procedure, finding significant differences between the near misses and near winners (Fig. S13ef). Moreover, to rule out the possibility that near misses have to work in the lab of established colleagues' after missing the funding, we also focused on last-author publications only[45], uncovering the same results (Fig. S20).

We also tested whether there exists any effect of near miss on team size, the number of different affiliations, as well as the probability to publish first-author/middle-author/last-author articles in the next 10 years after treatment, finding insignificant results for all cases (Fig. S21, $p$-value > 0.1). Finally, we tested if near misses teamed up with higher-impact collaborators by calculating the highest h-index among collaborators for each publication. We find no support for the hypothesis that near misses worked with higher impact collaborators in the 10 years after treatment (Fig. S21, $p$-value > 0.1). Together, these results indicate that the observed effect is unlikely to be attributable to collaborations.

## S4.3 Hypothesis *C*: Research direction shifts

Near misses may shift their research directions while near winners continue in the direction that they initially proposed. Were near misses more likely to change their research focus to a different field, or perhaps a hot one that tends to garner higher impact? To test whether there is any research direction shift for near misses within 10 years after treatment, we estimated the effect of early-career setbacks on research directions. To quantify the research direction of a certain publication, we measured keywords overlap between each article published after "treatment" and the set of articles published *ex ante*. Here, we consider MeSH Headings from the PubMed dataset as keywords, and measure their similarities using the Jaccard index. Comparing publications before and after the treatment, we find little evidence that early-career setbacks affected the research direction, which is consistent with prior studies[29] ($p$-value > 0.7, Fig. S21).

To test if the near misses may be more likely to publish on hot topics, we estimated the effect of early-career setbacks on the probability to publish an article on hot topics. We quantify hot-topic papers as papers with the most frequently occurring MeSH terms (top 5%) across all papers in the same year[46]. If a certain fraction of a paper's MeSH Headings belong among those that most frequently occurred, the paper may be considered as a hot-topic paper. We tried several variations based on specific fractions and ran a 2SLS regression for each case, where the dependent variable is the probability of publishing a paper on such hot topics.

First, we define a hot-topic paper as a paper where at least one of its MeSH terms belongs to the most frequently occurring MeSH terms. Under this definition, we find no significant effect of early-career setbacks on the likelihood to publish hot-topic papers in the future (coeff.=-0.0009, $p$-value=0.37). Second, we define a hot-topic paper as one where the vast majority (over 90%) of its MeSH terms all belong to the most frequently occurring MeSH terms. We find again no significant effect (coeff.=2%, $p$-value=0.12, Fig. S21). We did observe some suggestive evidence when we define a hot-topic paper as half of its MeSH terms (50%) belonging to the most frequently occurring category (coeff.=4.2%, $p$-value=0.067), but the effect weakened following the conservative removal procedure (coeff.=3.9%, $p$-value=0.101). Note that, near misses and near winners have the same probability to publish papers with hot topics *ex-ante* ($p$-value > 0.1). Finally, we repeated the 2SLS regression on publishing hit papers when controlling for hot topic papers, finding that, irrespective to the varied definitions, the result remains significant ($p$-value = 0.018 for the conservative removal), indicating that changing research direction alone cannot account for the main effect.

## S4.4 Hypothesis D: Changing institutions

Near misses may change research institutions with increased frequency following early-career setbacks, and such moves might expose these scientists to new sets of ideas or new collaborators. To test this hypothesis, we trace the physical mobility of each PI through their affiliations recorded in R01 applications, and calculate the probability of changing institutions in the ten-year time window. We find that near misses had a 40% chance to change their initial institution, whereas near winners had a 42% chance to move in the next ten-year window ($p$-value = 0.515). 2SLS regression yielded similar results ($p$-value = 0.235), suggesting that physical movements are unlikely to be the source of the observed performance gap.

## S4.5 Combining hypotheses *A - D*

To test if there is combining all above hypothesis explain our main findings, we controlled all mentioned processes and ran a 2SLS regression, as follows:

*1$^{st}$ stage:* $F_j = \alpha_0 + \alpha_1 s_j + \alpha_2 s_j^2 + \cdots + \alpha_p s_j^p + \pi_i z_j + \theta X_{i,pre} + \delta X_{i,attr} + \mu_t + \eta_n + \eta_j,$

*2$^{nd}$ stage:* $y_{it} = \beta_0 + \beta_1 s_j + \beta_2 s_j^2 + \cdots + \beta_p s_j^p + \lambda \hat{F}_j + \gamma X_{i,pre} + \delta' X_{i,attr} + \mu_t' + \eta_n' + \epsilon_i,$

where $X_{i,attr}$ is the above mentioned hypothesis, i.e., novelty, team size, number of different affiliations, hot topic, author order, MeSH Heading overlapping with prior publications, etc. Controlling all these parameters, we find our main effects remained, suggesting additional processes may be at work (coef. = 7.1% *p*-value = 0.012 for active PIs; coef. = 6.6% *p*-value = 0.021 for the conservative removal).

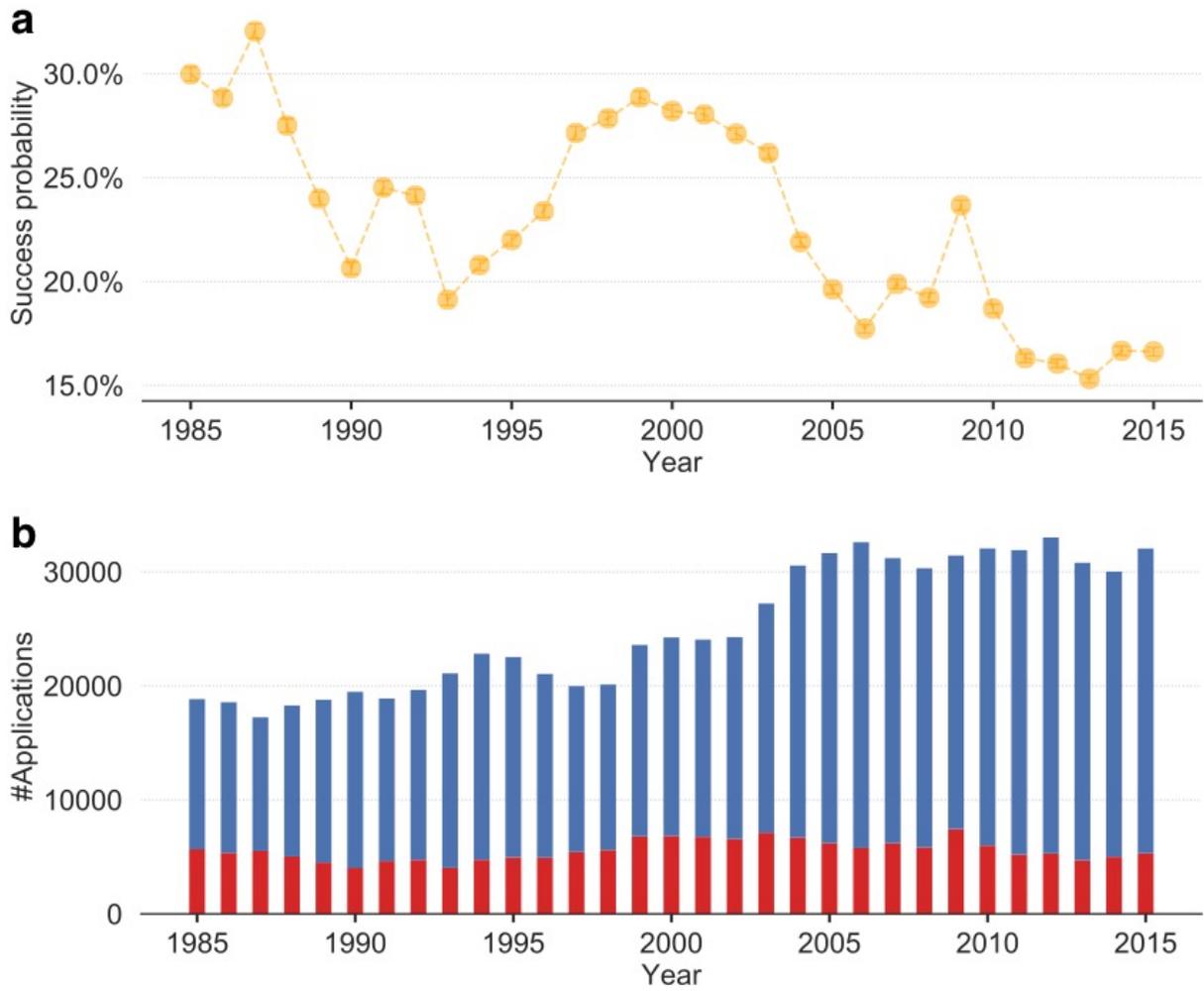

**Figure S1: NIH funding landscape. (a)** R01 success probability as a function of time. **(b)** number of successful (red) and all R01 competing applications (blue) as a function of time.

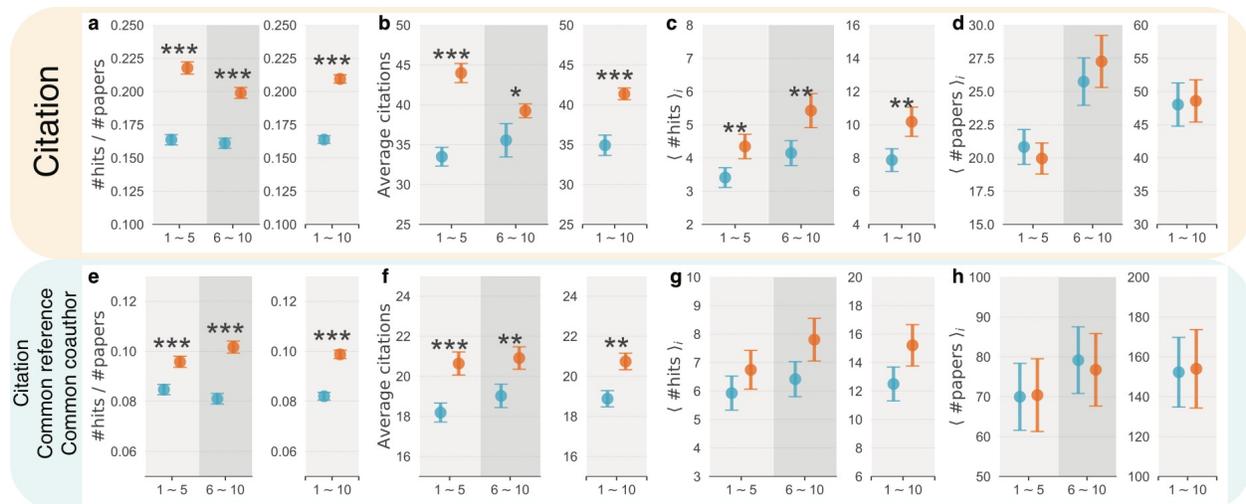

**Figure S2: Comparing near winners and near misses using different name disambiguation methods.** The comparison of hit rate per paper (**a**), average citations within 5 years of publication (**b**), average number of hit papers (**c**) and average number of papers (**d**) between near winners and near misses. Here, if two papers from the initial pool are considered from the same author if they cite each other. (**e - h**) The same as (**a - d**), and two papers from the initial pool are considered from the same author if one of the following conditions is fulfilled: they cite each other, or share at least one common reference, or share at least one common coauthor name. When calculating the average citations, we use the NIH applications from 1990 to 2000. *** $p < .001$, ** $p < .05$, * $p < .1$; Error bars represent the standard error of the mean.

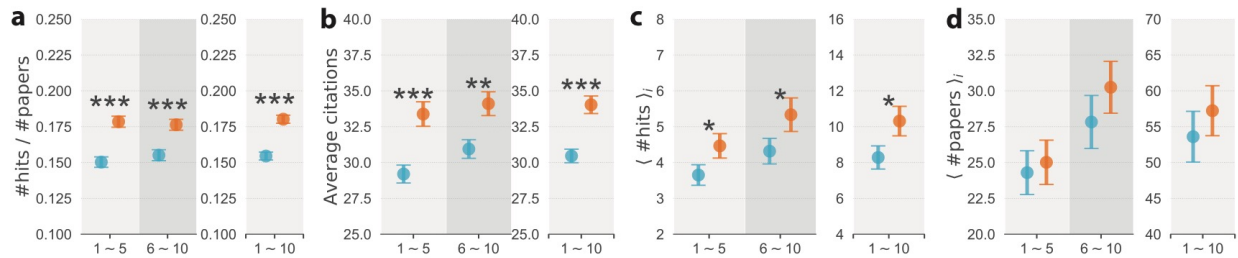

**Figure S3: Comparing near misses and near winners after removing PIs with common Asian surnames. (a)** hit rate per paper; **(b)** average citations within 5 years of publications, here we focus on NIH grant applications from 1990 to 2000; **(c)** number of hit papers per person; **(d)** number of publications per person. *** p < .001, ** p < .05, * p < .1; Error bars represent the standard error of the mean.

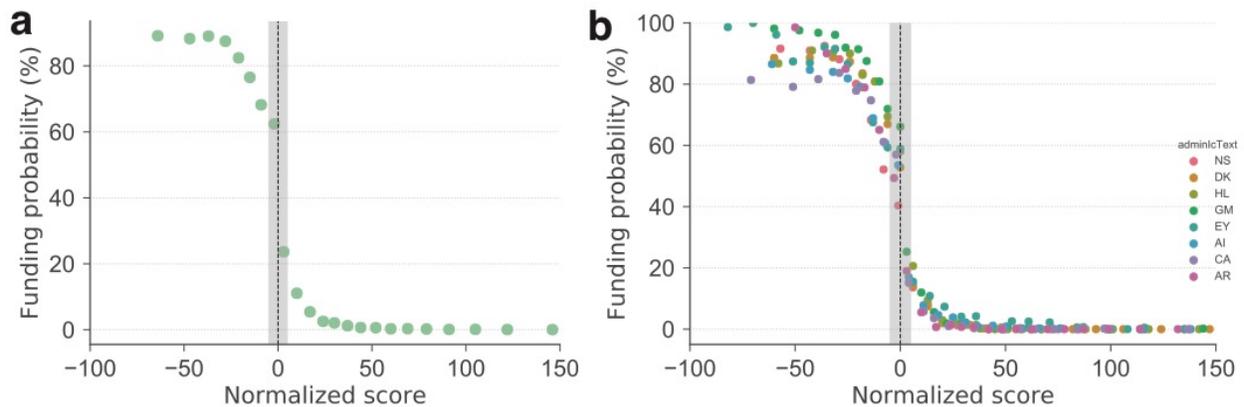

**Figure S4: Relationship between funding success probability as a function of score. (a)** For all NIH grant applications considered in our study, the funding probability as a function of normalized score; **(b)** For major NIH institutes, which includes more than 75% of applications, the funding probability as a function of normalized score.

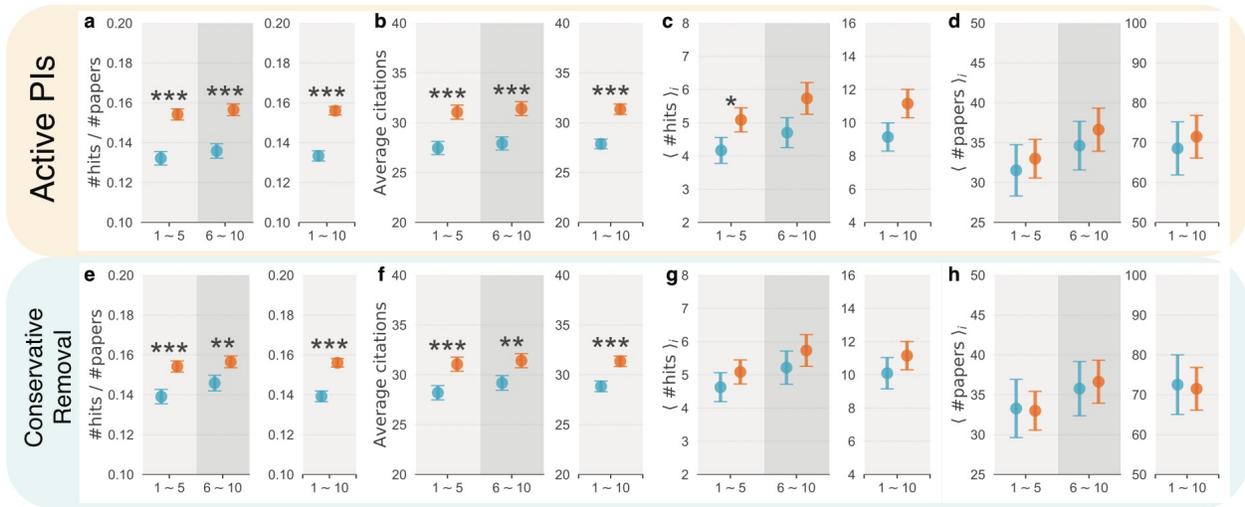

**Figure S5: Comparing near misses and near winners using cutoff the second worst funded applications in each study section.** **(a)** hit rate per paper; **(b)** average citations within 5 years of publications, here we focus on NIH grant applications from 1990 to 2000; **(c)** number of hit papers per person; **(d)** number of publications per person. **(e - h)** The same as **a - d** but using the conservative removal method. *** p < .001, ** p < .05, * p < .1; Error bars represent the standard error of the mean.

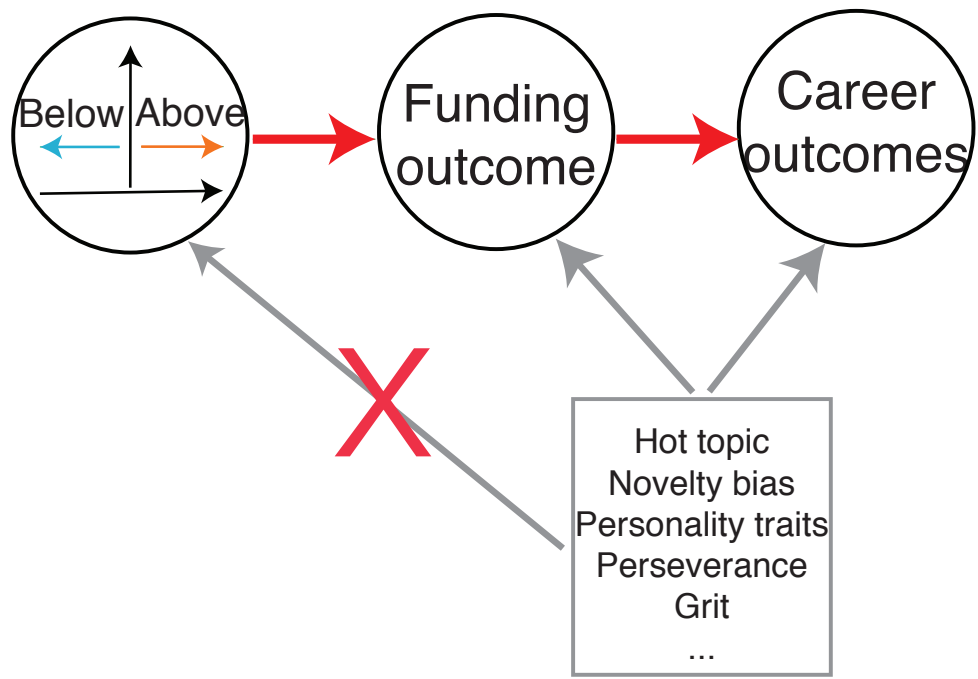

**Figure S6 An illustration of the fuzzy RD approach**. In this figure, above or below the funding cutoff is treated as an instrumental variable (IV). The IV framework helps us disentangle unobserved factors by using variation created by the IV as an exogenous shock to one endogenous variable to estimate its causal effect on another variable. For example, there could be unobserved factors or individual characteristics that might influence both the funding and career outcomes (gray box and arrows), but these hidden variables differ smoothly with the score and are uncorrelated with the IV (gray arrow with a red cross). Whether or not one's score is above the threshold (the IV) only affects the funding outcome, but is uncorrelated with future career outcomes. Hence if the IV itself predicts future career outcomes, it would mean that the pathway indicated by red arrows must operate[47,48], allowing us to further establish a causal link between early-career near miss and longer-term scientific success.

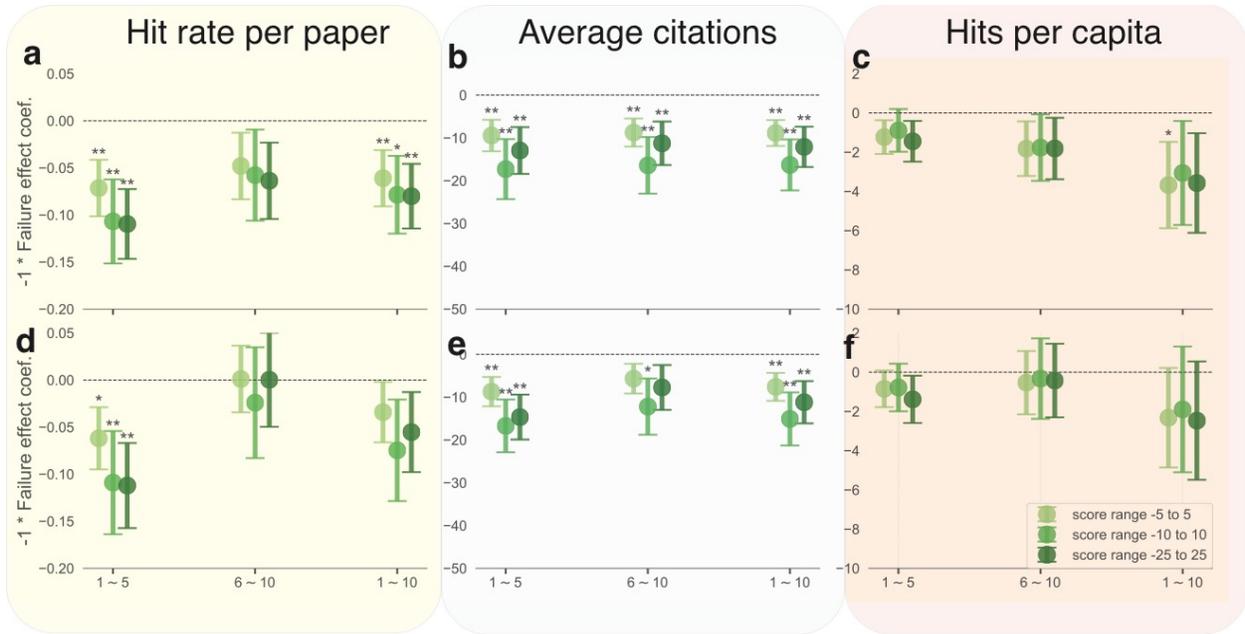

**Figure S7: The estimation from the fuzzy RDD estimation. (a)** The effect of near miss on the probability to publish top 5% hit papers for applicants in the NIH system; **(b)** the effect of near miss on the average citations within 5 years after publication for applications in the NIH system; Here we use data from 1990 to 2000. **(c)** the effect of near miss on the number of top 5% hit papers published for applicants in the NIH system; **d - f** The same as **a - c** but for the conservative removal. Here, we use three different sample size, i.e., 5-score around the cutoff, 10-score from the cutoff, and 25-score from the cutoff. We add exclusive including NIH institution and time fixed effect, and PI prior performance (see Supplementary Information S3). Error bars represents the standard errors, and are clustered at individual level. *** $p < .001$, ** $p < .05$, *$p < .1$.

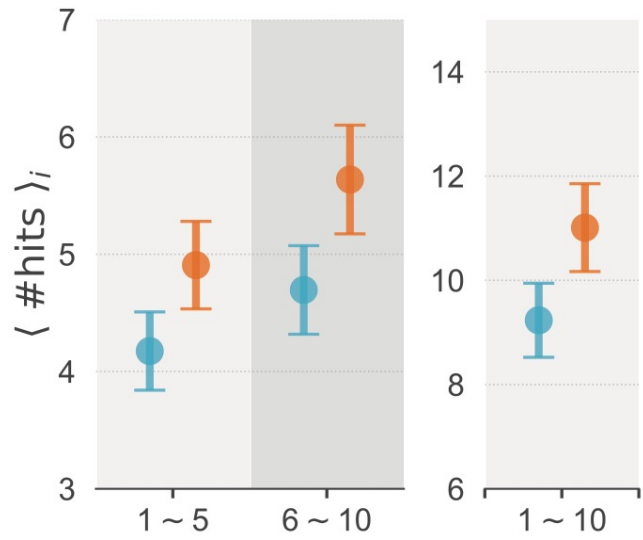

**Figure S8: Comparing future career outcome between near misses (orange) and near winners (blue) in terms of hits per capita.** Hits per capita measure the number of hit papers per person. We uncovered the same direction of results as Figure 2 in the main text, albeit with a lesser significance level due to a reduced sample size (p-value = 0.107). Error bars represent the standard error of the mean.

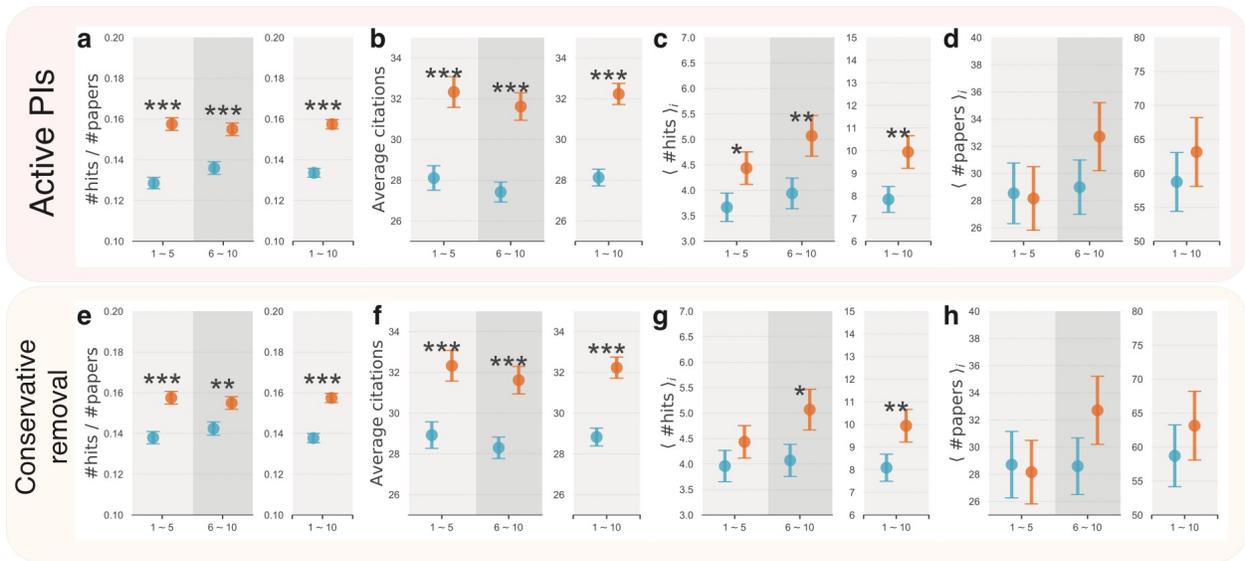

**Figure S9: Comparing near misses with near winners for those who applies the first R01 grant at the time of treatment. a - d,** The comparison of hit rate per paper, average citations within 5 years after publication (using data from 1990 to 2000), hits per capita and number of publications per capita for active PIs; **e - h,** The same as **a - d** but for the conservative removal. *** $p < .001$, ** $p < .05$, *$p < .1$; Error bar represents the standard error of the mean.

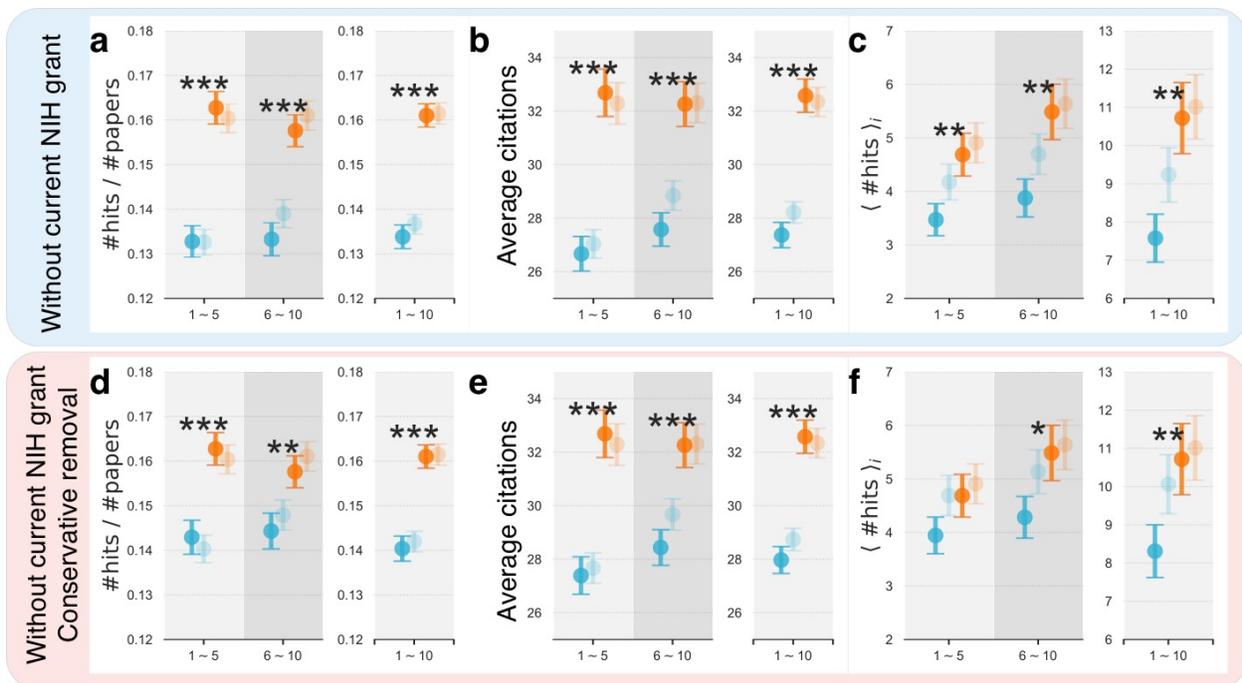

**Figure S10: The effect of a near miss magnifies when the stake gets higher. a - c,** The subsequent performance by the near-miss and near-win groups, measured by the probability of producing a hit paper (**a**), average citations of papers attained within 5 years of publication; here we used data from 1990 to 2000 (**b**), and hit papers per capita (**c**). Shaded symbols represent the corresponding measurements reported in Fig. 2 of the main text. **d - f**, The same as **a - c** but for the conservative removal. Shaded symbols represent the corresponding measurements reported in Fig. 3 of the main text. *** $p < .001$, ** $p < .05$, *$p < .1$. Error bar represents the standard error of the mean.

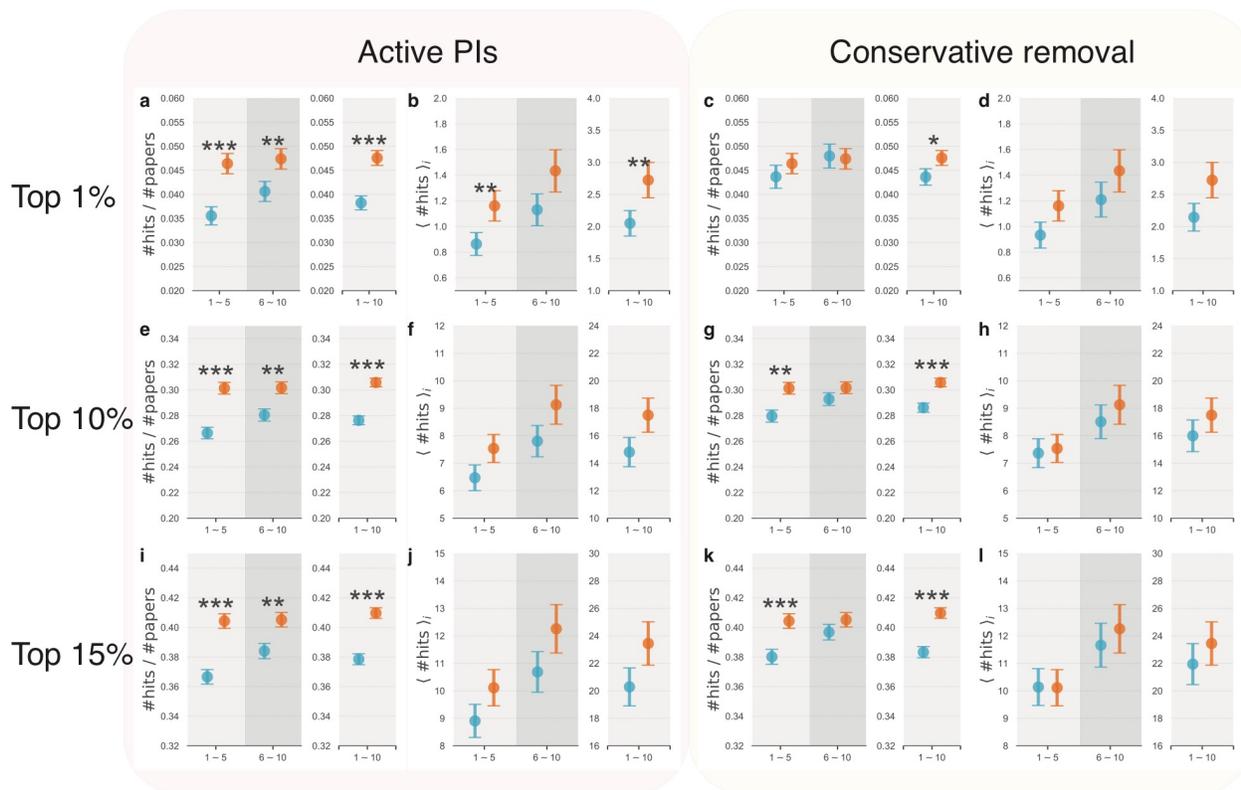

**Figure S11: Robustness check for various hit paper thresholds.** Comparison between near winners and near misses for different thresholds of hit papers. **a-d,** top 1% highly cited publications; **e-h,** top 10% highly cited publications; **i-l,** top 15% highly cited publications. \*\*\* p < .001, \*\* p < .05, \*p < .1; Error bar represents the standard error of the mean.

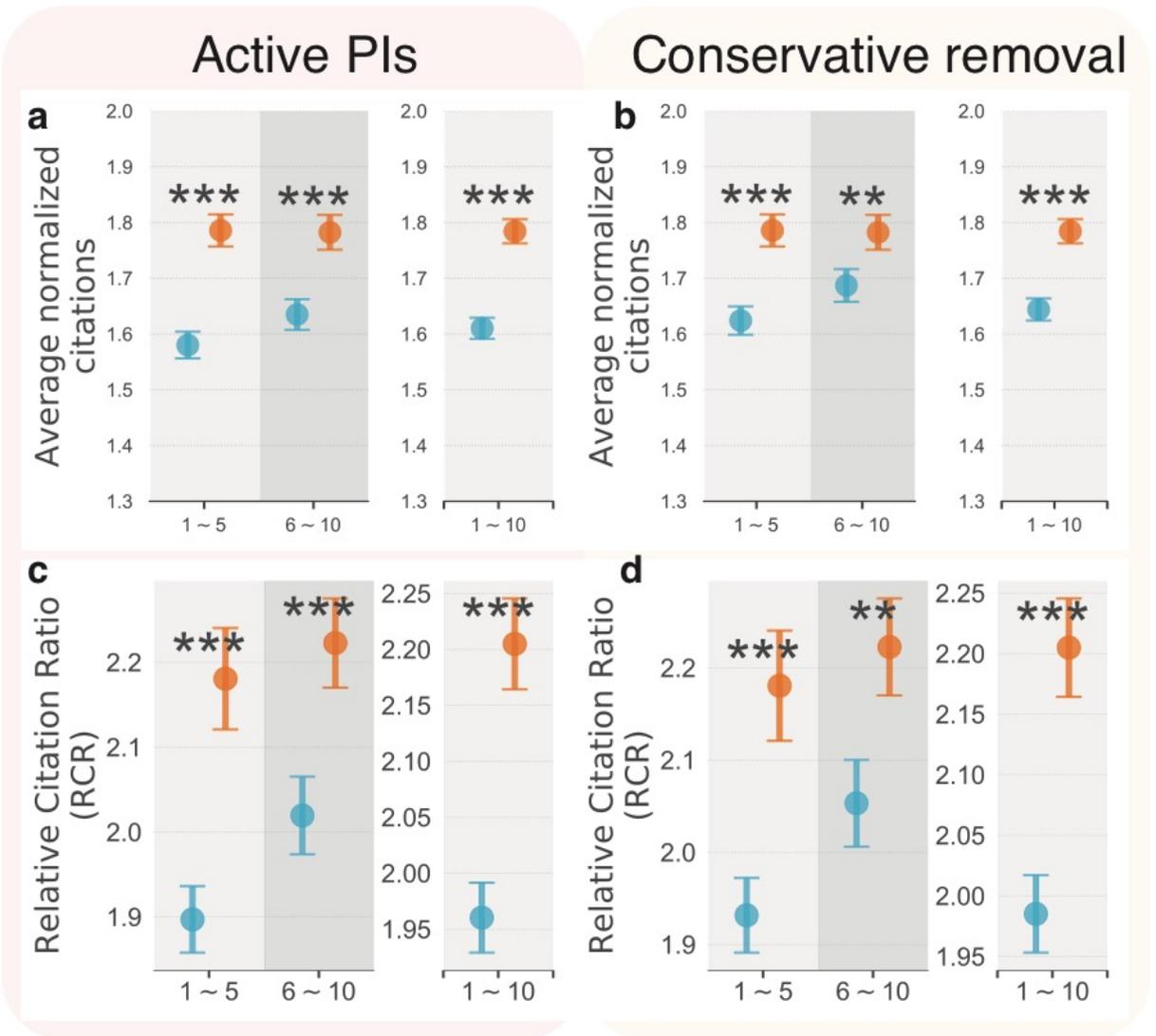

**Figure S12: Robustness check for the normalized citations with respect to field and time.** Comparing normalized citations $c_f$ between near winners and near misses for **a,** active PIs and **b,** conservative removal. Comparing the relative citation ratio (RCR) between near winners and near misses for **c**, active PIs and **d**, conservative removal. *** p < .001, ** p < .05, *p < .1; Error bar represents the standard error of the mean.

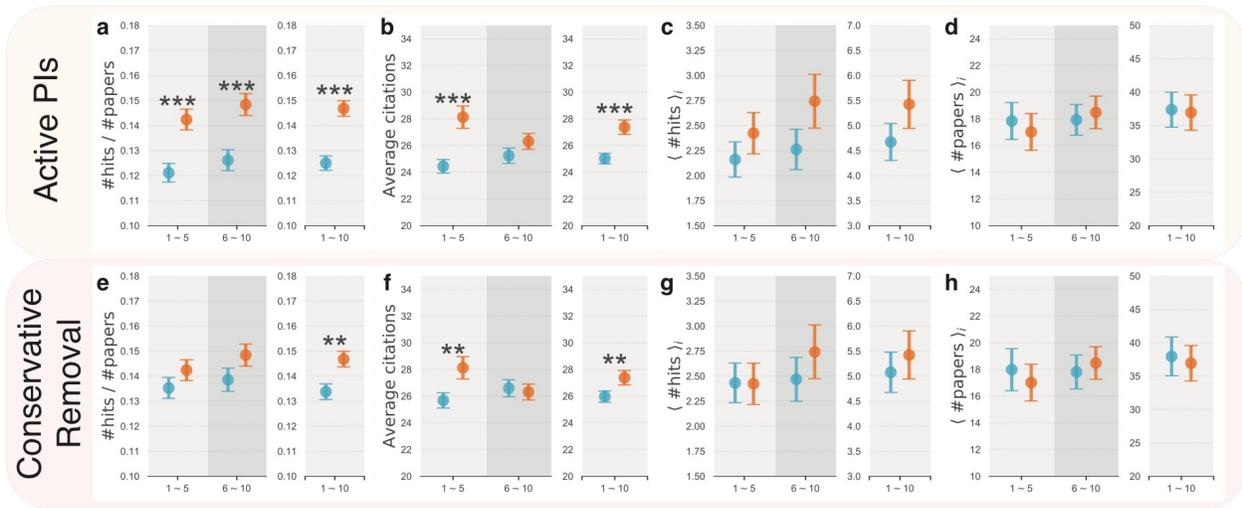

**Figure S13: Robustness check for lead-author publications.** **a-d,** Comparing near winners with near misses in terms of hit rate per paper, average citations within 5 years after publication (using data from 1990 to 2000), hits per capita, and number of publications per person for active PIs; **e-h,** The same as **a-d** but we use the conservative removal method to account for the screening effect. *** $p < .001$, ** $p < .05$, *$p < .1$; Error bar represents the standard error of the mean.

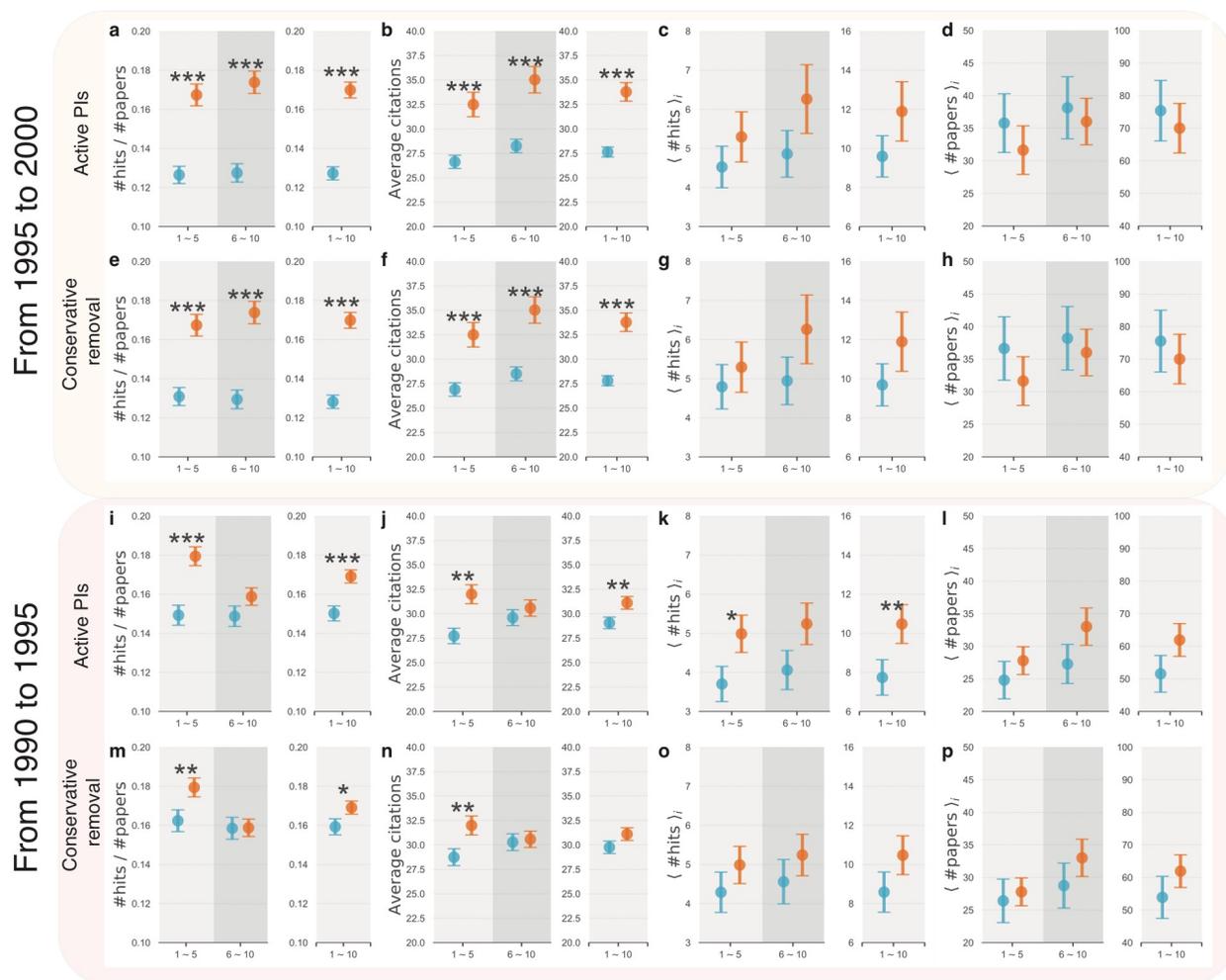

**Figure S14: Robustness check for different time windows. a-d,** The comparison between near misses and near winners in terms of hit rate per paper, average citations per paper after five years of publication, hits per capita, number of publications per person for top 5% paper between 1995 to 2000 for active PIs; **e-h,** The same as **a-d** but for the conservative removal. **i-l,** The comparison between near misses and near winners in terms of hit rate per paper, average citations per paper after five years of publication, hits per capita, number of publications per person for top 5% paper between 1990 to 1995 for active PIs; **m-p,** The same as **i-l** but for the conservative removal. *** $p < .001$, ** $p < .05$, *$p < .1$; Error bar represents the standard error of the mean.

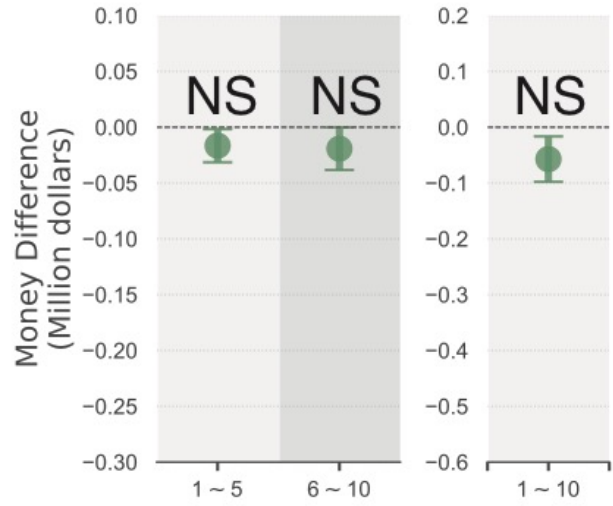

**Figure S15: Comparing the NSF money between near misses and near winners.** *** p < .001, ** p < .05, *p < .1 and NS for p > .1; Error bar represents the standard error of the mean.

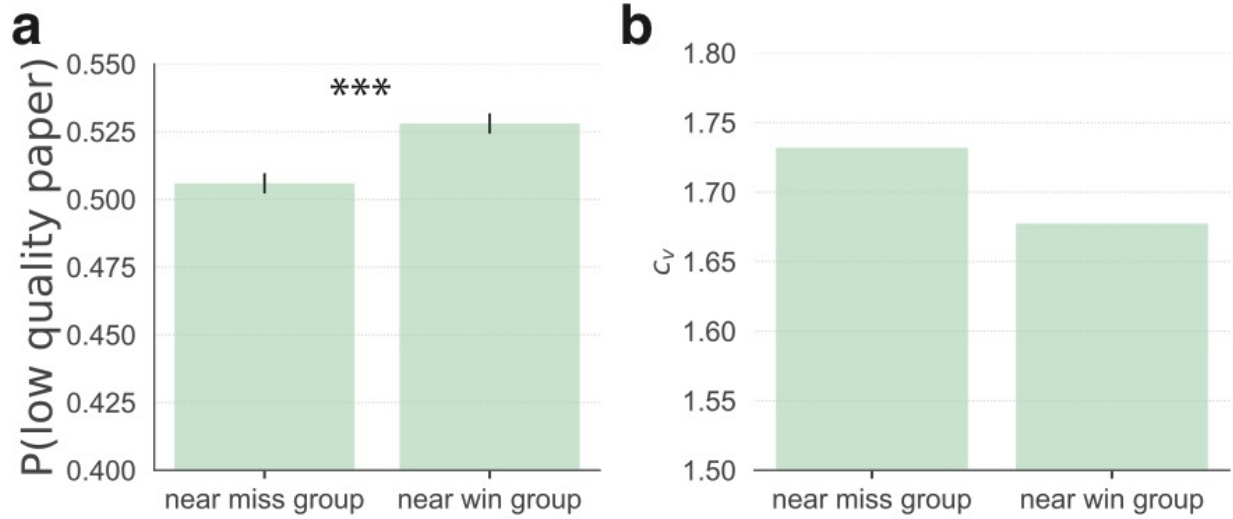

**Figure S16: Comparing citation uncertainty between near misses and near winners.** (a) Probability to publish low quality papers, with citations lower than average citations of papers published in the same field and time; Error bar represents the standard error of the mean. (b) Coefficient of variance of $c_f$ for both groups. *** $p < .001$, ** $p < .05$, *$p < .1$.

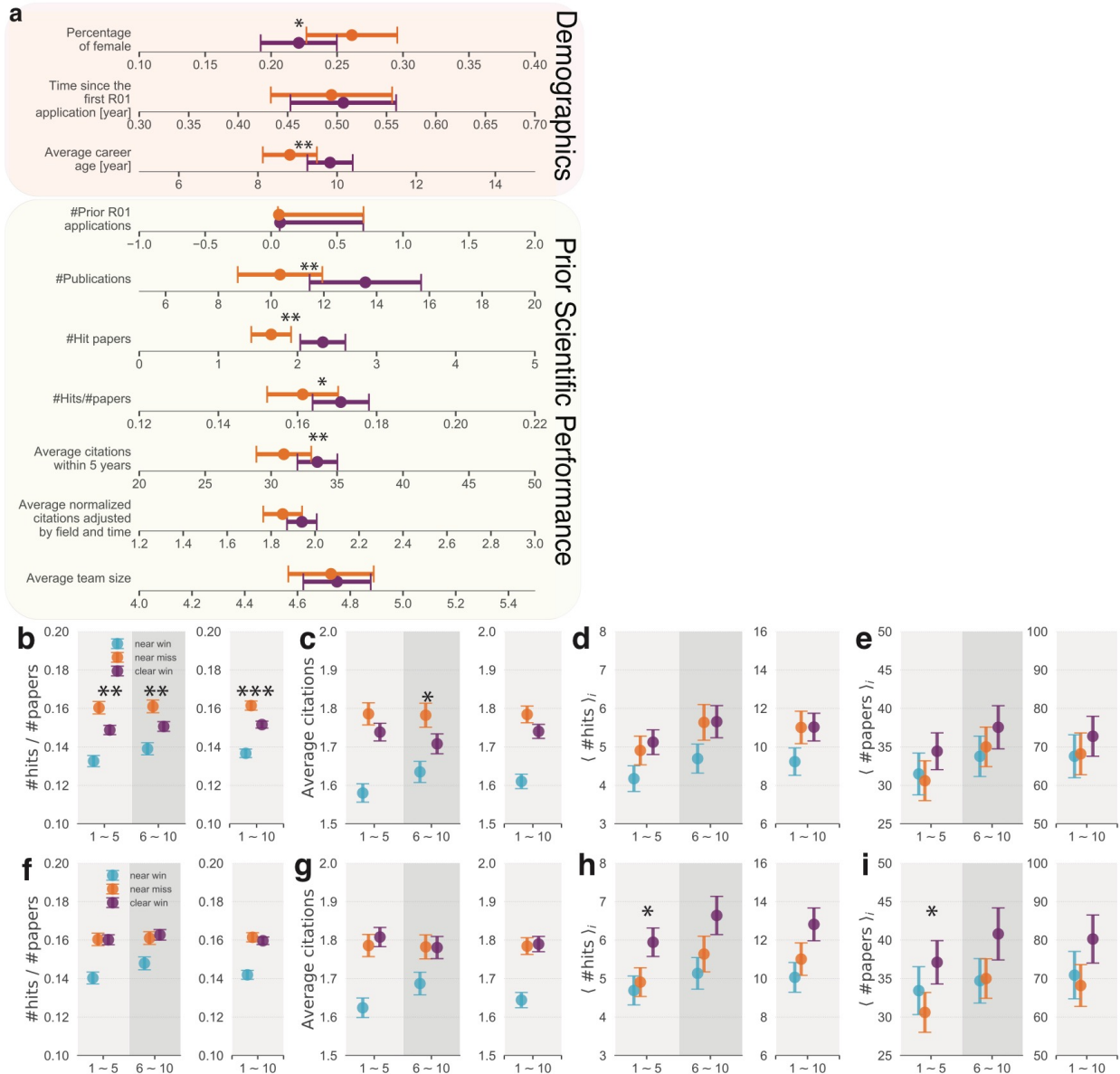

**Figure S17: Comparing near misses with clear winners in terms of various measurements, with the basis of near winners.** **(a)** Comparing features of near misses with clear winners prior to treatment; detailed definitions can be found at Fig. 1 of the main text. Error bar represents the 95% confidence interval; **b-e,** Comparing hit rate per paper (**b**), average normalized citations (**c**), number of hit papers per capita (**d**), number of papers (**e**); **f-i,** The same as **b-e**, but for the conservative removal. *** $p < .001$, ** $p < .05$, *$p < .1$; Error bar represents the standard error of the mean.

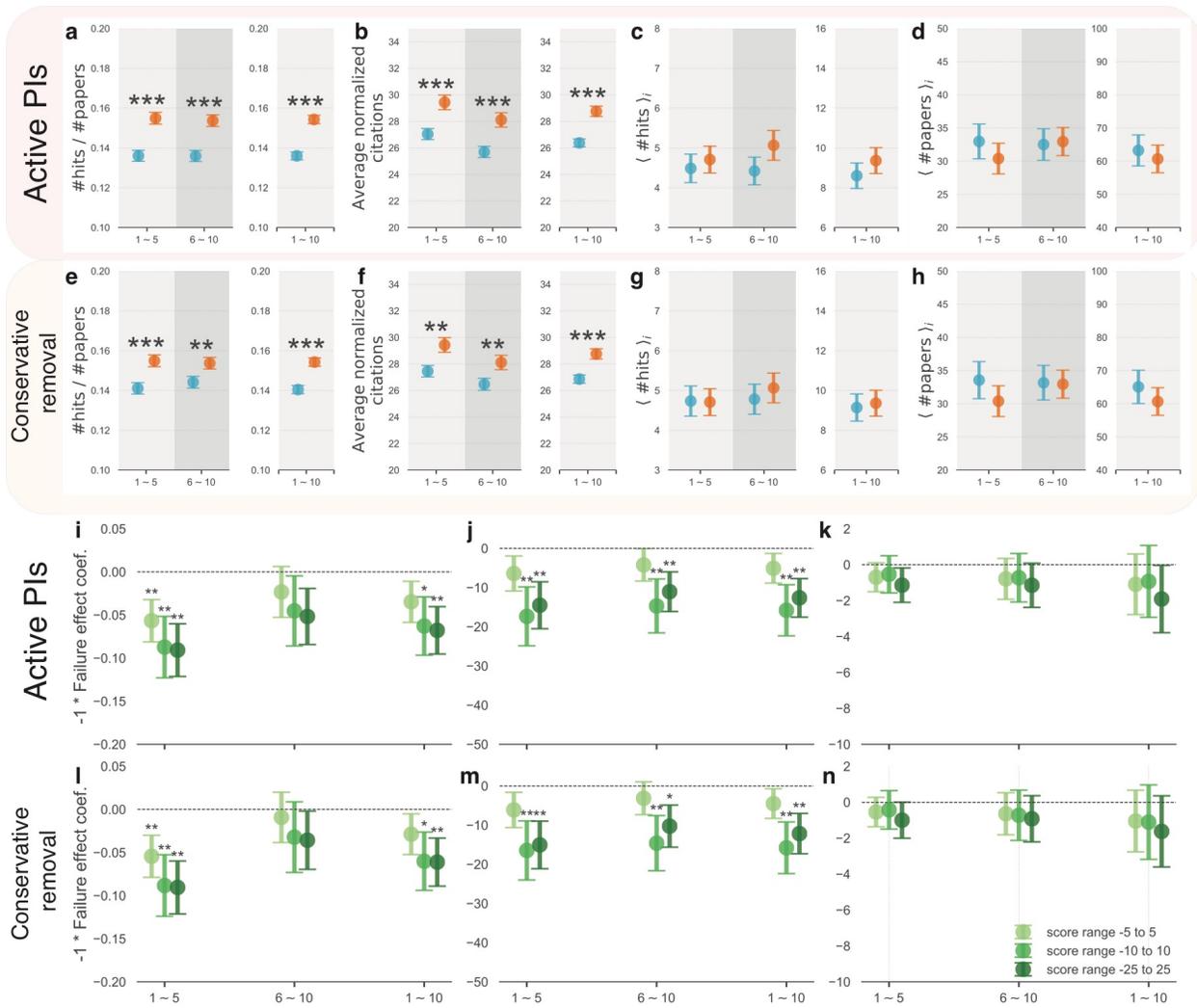

Figure S18: Comparing near misses with near winners, here the active PI is defined as those who publish at least one paper during certain time windows. **a-d,** The comparison between near misses and near winners in terms of hit rate per paper, average citations within 5 years after publication (using data from 1990 to 2000), hits per capita and papers per capita by focusing only on active PIs; **e-h,** The same as **a-d** but for conservative removal; Error bar represents the standard error of the mean. **i-k,** fuzzy RDD estimation of active PIs; **l-n,** fuzzy RD estimation of PIs after the conservative removal. In the regression estimation, error bars are standard errors clustered at individual level. *** $p < .001$, ** $p < .05$, * $p < .1$.

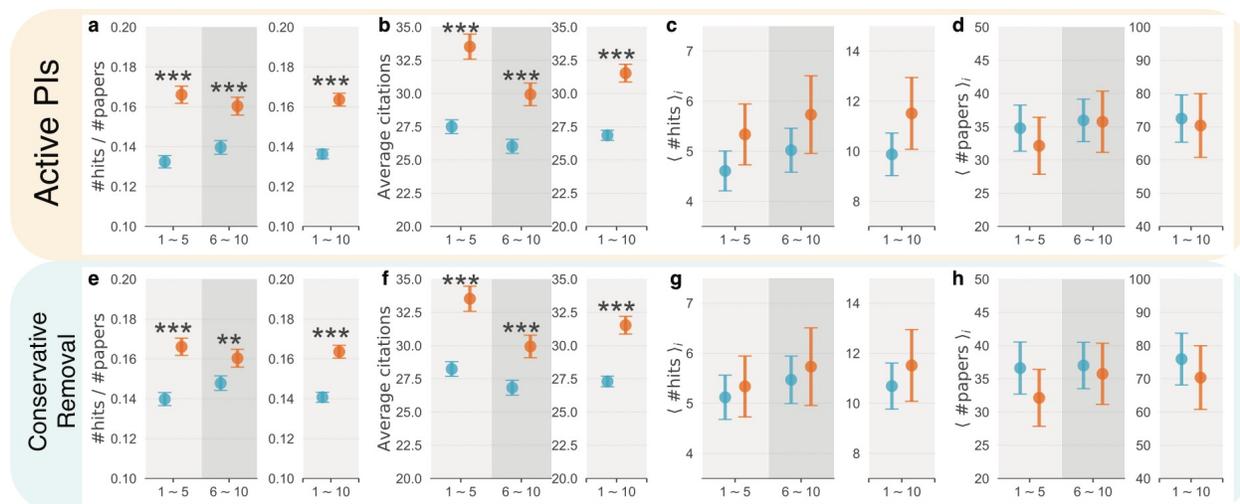

**Figure S19: Comparing near misses with near winners using cutoff defined by the percentile score. a-d,** Comparing near winners with near misses in terms of hit rate per paper, average citations within 5 years after publication (using data from 1990 to 2000), hits per capita, and number of publications per person for active PIs; **e-h,** The same as **a-d** but we use the conservative removal method to account for the screening effect. *** $p < .001$, ** $p < .05$, *$p < .1$; Error bar represents the standard error of the mean.

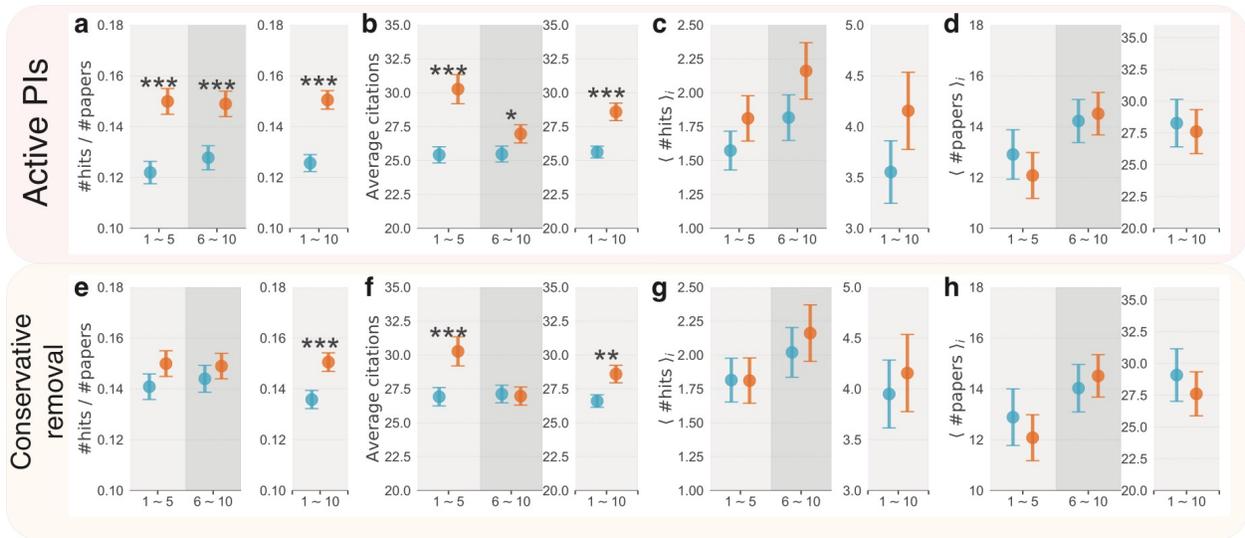

**Figure S20: Results for last-author publications only**. **a-d,** Comparing near winners with near misses in terms of hit rate per paper, average citations within 5 years after publication (using data from 1990 to 2000), hits per capita, and number of publications per person for active PIs; **e-h,** The same as **a-d** but we use the conservative removal method to account for the screening effect. *** p < .001, ** p < .05, *p < .1; Error bar represents the standard error of the mean.

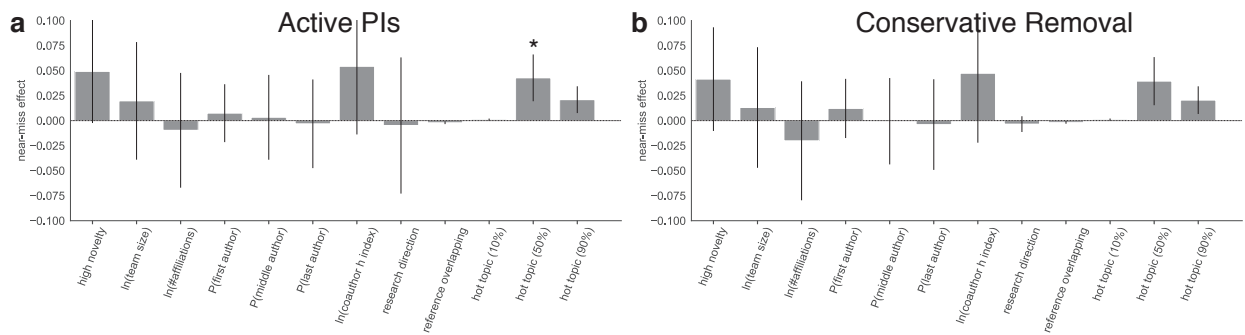

**Figure S21: Possible generative mechanisms of why near misses become better than near winners**. **(a)** We present regression estimates, studying the effect of early-career setbacks on ten possible mechanisms for active PIs: publishing high novel papers, team size, number of different affiliations, probability to publish first-author, middle-author, and last author publications, maximum coauthor h index of each publication, research direction, reference overlapping, and hot topic (we define a paper covers a hot topic if a certain fraction of its MeSH terms all belong to the most frequently occurring MeSH terms, i.e., 10%, 50%, and 90%.). **(b)** The same as A but for the conservative removal. *** p < .001, ** p < .05, *p < .1; Error bars represent the standard errors, and are clustered at individual level.

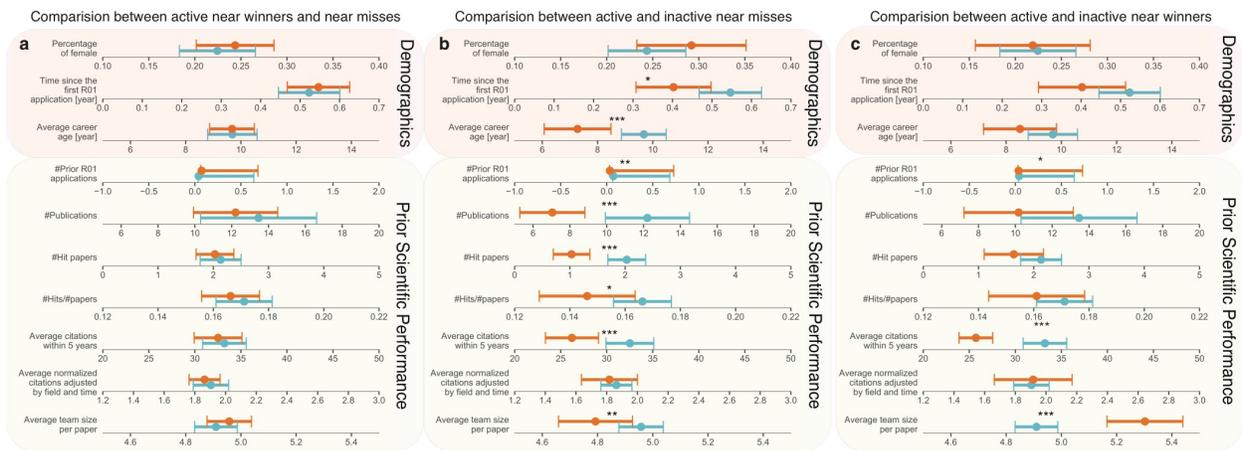

**Figure S22: Pre-treatment comparisons between the near-win and near-miss applicants.** (**a**) Pre-treatment feature comparisons between the near-miss and near-win group who are active in the next 10 years after treatment. We compared 10 different demographic and performance characteristics. The features are defined as follows (from top to bottom): 1) percentage of female applicants; 2) number of years since the first R01 application; 3) number of years since the first publication; 4) number of previous R01 applications; 5) number of publications prior to treatment; 6) number of prior papers that landed within the top 5% of citations within the same field and year; 7) probability of publishing a hit paper; 8) average citations papers received within 5 years of publication; 9) citations normalized by field and time11; and 10) average team size across prior papers. We see no significant difference between the two groups across any of the ten dimensions we measured. (**b**) The same as **a** but we compare active and inactive near misses. (**c**) The same as **a** but the comparison between active and inactive near winners. Error bar represents the 95% confidence interval. Blue and orange indicate near-win and near-miss applicants, respectively. *** $p < .001$, ** $p < .05$, * $p < .1$.

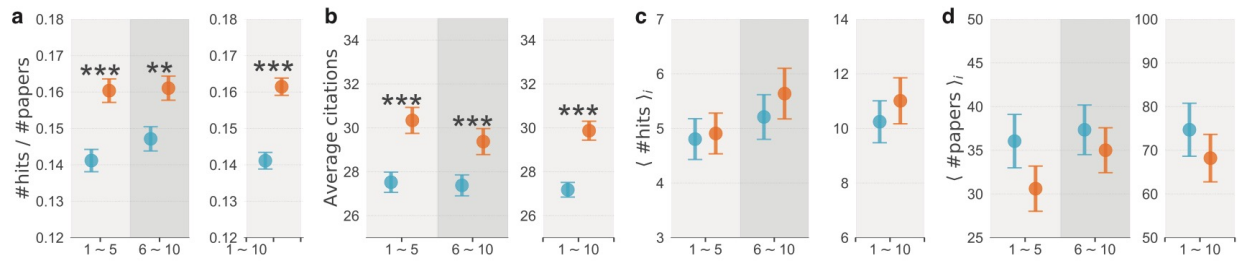

**Figure S23 Alternative ways to define less able PIs.** Comparing near winners with near misses in terms of (**a**) hit rate per paper, (**b**) average citations within 5 years after publication (using data from 1990 to 2000), (**c**) hits per capita, and (**d**) number of publications per person. *** $p < .001$, ** $p < .05$, * $p < .1$; Error bar represents the standard error of the mean.